# Large Language Model–driven Analysis of General Coordinates Network (GCN) Circulars

Vidushi Sharma[1,2], Ronit Agarwala[1,3], Judith L. Racusin[1], Leo P. Singer[1,4], Tyler Barna[5], Eric Burns[6], Michael W. Coughlin[5], Dakota Dutko[1,7], Courey Elliott[6], Rahul Gupta[1,11], Ashish Mahabal[8], and Nikhil Mukund[9,10]

[1] Astrophysics Science Division, NASA Goddard Space Flight Center, Greenbelt, MD 20771, USA; vidushi.sharma@nasa.gov, ragarwala@ucsd.edu
[2] Center for Space Science and Technology, University of Maryland Baltimore County, Baltimore, MD 21250, USA
[3] University of California San Diego, La Jolla, CA 92093, USA
[4] Joint Space-Science Institute, University of Maryland, College Park, MD 20742, USA
[5] School of Physics and Astronomy, University of Minnesota, Minneapolis, MN 55455, USA
[6] Department of Physics & Astronomy, Louisiana State University, Baton Rouge, LA, USA
[7] ADNET Systems Inc., MD, USA
[8] Center for Data-Driven Discovery, California Institute of Technology, Pasadena, CA 91125, USA
[9] MIT Kavli Institute for Astrophysics and Space Research and LIGO Laboratory, Massachusetts Institute of Technology, Cambridge, MA 02139, USA
[10] NSF AI Institute for Artificial Intelligence and Fundamental Interactions (IAIFI), Cambridge, MA, USA
*Received 2025 October 20; revised 2025 November 17; accepted 2025 November 18; published 2026 February 27*

## Abstract

The General Coordinates Network (GCN) is NASA's time-domain and multimessenger alert system. GCN distributes two data products: automated "Notices" and human-generated "Circulars" that report the observations of high-energy and multimessenger astronomical transients. The flexible and nonstructured format of GCN Circulars, comprising more than 40,500 Circulars accumulated over three decades, makes it challenging to manually extract observational information, such as redshift or observed wave bands. In this work, we employ large language models (LLMs) to facilitate the automated parsing of transient reports. We develop a neural topic modeling pipeline with open-source tools for the automatic clustering and summarization of astrophysical topics in the Circulars archive. Using neural topic modeling and contrastive fine-tuning, we classify Circulars based on their observation wave bands and messengers. Additionally, we separate gravitational-wave event clusters and their electromagnetic counterparts from the Circulars archive. Finally, using the open-source `Mistral` model, we implement a system to automatically extract gamma-ray burst (GRB) redshift information from the Circulars archive, without the need for any training. Evaluation against the manually curated Neil Gehrels Swift Observatory GRB table shows that our simple system, with the help of prompt-tuning, output parsing, and retrieval augmented generation (RAG), can achieve an accuracy of 97.2% for redshift-containing Circulars. Our neural search-enhanced RAG pipeline accurately retrieved 96.8% of redshift Circulars from the manually curated archive. Our study demonstrates the potential of LLMs to automate and enhance astronomical text mining and provides a foundational work for future advances in transient alert analysis.

## 1. Introduction

Time-domain and multimessenger astronomy have advanced significantly in the past decade, particularly with the detection of gravitational waves (GWs; B. P. Abbott et al. 2017, 2018). This breakthrough has led to an increase in observational reports, with the astronomy community showing magnified interest in following up on high-energy transients, high-energy neutrinos, and GWs. Additionally, new instruments dedicated to the study of the most energetic phenomena in the Universe (e.g., supernovae, GRBs, and other bursts of Galactic and extragalactic nature) are contributing to the General Coordinates Network (GCN). To keep pace with technological advancements and community needs, the new GCN was launched in 2022 July (J. L. Racusin et al., in preparation). With upgraded technology and increased interest in multimessenger astronomy, the GCN is a reliable and secure platform. There will be a huge amount of data in the future that will require automated methods for analysis. Even new transients detected with the Space Variable Objects Monitor (J. Paul et al. 2011) and the Einstein Probe (W. Yuan et al. 2025) have reported an increase in GRBs and new X-ray transients, driving the rise in follow-up observations from the community. Moreover, observatories reporting fast radio bursts (FRBs), such as the Canadian Hydrogen Intensity Mapping Experiment (CHIME; CHIME/FRB Collaboration et al. 2018) and Deep Synoptic Array-110 (C. J. Law et al. 2024), will join the network soon. These efforts will be complemented by significant advancements and commissioning of new observatories designed to detect transients, such as high-energy neutrinos (e.g., IceCube; M. G. Aartsen et al. 2017, KM3NeT; KM3NeT Collaboration et al. 2024), GWs

[11] NASA Postdoctoral Program Fellow.

(e.g., Virgo, Laser Interferometer Gravitational-Wave Observatory (LIGO), Kamioka Gravitational Wave Detector (KAGRA), LISA, radio (e.g., Square Kilometre Array; P. E. Dewdney et al. 2009), optical observations (e.g., Large Synoptic Survey Telescope; Ž. Ivezić et al. 2019), and high-energy gamma rays (e.g., Cherenkov Telescope Array (Cherenkov Telescope Array Consortium et al. 2019). Therefore, a wealth of information will be available in both multiwavelength and multimessenger astrophysics, with increasing interest in following up on exciting events.

GCN is a public collaboration platform managed by NASA for the astronomy research community. It enables the sharing of real-time alerts and rapid communications regarding high-energy, multimessenger, and transient phenomena. GCN distributes alerts for both space and ground-based observatories, physics experiments, and thousands of astronomers worldwide. Established in 1992, GCN began as an innovative solution for near-time notification of gamma-ray bursts (GRBs) detected by the Burst and Transient Source Experiment on board the Compton Gamma Ray Observatory (S. D. Barthelmy et al. 1994, 1995). In 1997, as new instruments and missions were added to the network, it was renamed the Gamma-ray burst Coordinates Network. GCN has enabled global follow-up observations that have significantly improved our understanding of GRBs, including their afterglows, redshifts (e.g., first redshift for GRB 970508; D. E. Reichart 1998), supernova detections (e.g., SN 1998bw/GRB 980425; T. J. Galama et al. 1998), and host-galaxy information. Over time, the network expanded beyond the GRB community to include other high-energy astronomical transients, such as high-energy neutrino (e.g., IceCube-170922A; C. Kopper & E. Blaufuss 2017), soft gamma-ray repeaters (SGRs; e.g., GRB 200415A; A. Pozanenko et al. 2020), and tidal disruption events (TDEs; e.g., Swift J1644+57; D. N. Burrows et al. 2013).

GCN provides two types of alerts to support research in astrophysical transient phenomena: GCN Notices and GCN Circulars. GCN Notices are brief, machine-generated reports.[12] GCN Circulars,[13] on the other hand, are more detailed, human-written observational reports that often contain information about observations, predictions, requests for follow-ups, and plans for future observations for multimessenger astrophysical events. GCN Circulars (hereafter, Circulars) have a flexible and unstructured format. Over the past 30 yr, more than 40,500 Circulars have been submitted, making the Circulars archive a valuable resource for astrophysicists studying multimessenger events. However, this volume presents a challenge in managing the growing volume of astrophysical data and alerts to extract useful information, given the varying structures of Circulars.

GCN plays a central role as a global distribution system for alerts from high-energy and multimessenger observatories. Other brokers like Fink (A. Möller et al. 2021) or ALerCE (F. Förster et al. 2021) ingest alerts from time-domain surveys and provide value-added annotations, for example, through crossmatching with external catalogs. While brokers enrich survey data, GCN facilitates the rapid dissemination of alerts to the community and provides a standardized framework for real-time and archival access. This infrastructure enables swift coordination among observers and follow-up facilities, complementing broker-based alert streams and linking them to the broader multimessenger landscape. Transient objects identified by brokers are often shared with the community via the Transient Name Server,[14] the Minor Planet Center,[15] Astrophysical Multimessenger Observatory Network alerts (H. A. Ayala Solares et al. 2020), or GCN, depending on the nature of the source. Coordinated follow-up efforts are then facilitated by Target and Observation Managers (TOM), with notable examples including the Young Supernova Experiment (D. A. Coulter et al. 2023), the TOM Toolkit (R. A. Street et al. 2018), `SkyPortal` (S. J. van der Walt et al. 2019; M. W. Coughlin et al. 2023), and Astro-COLIBRI (P. Reichherzer et al. 2021).

Recently, there has been a growing interest in integrating natural language processing (NLP) tools into astrophysical research. For instance, F. Grezes et al. (2024) highlight the potential of astroBERT, a large language model (LLM) based on the Bidirectional Encoder Representations from Transformers (BERT) deep neural network architecture (J. Devlin et al. 2018). Trained on 15 million records from NASA ADS, astroBERT demonstrates capabilities in recognizing named entities within astrophysical texts. N. Mukund et al. (2018) show the application of NLP to retrieve and display relevant logbook data, aiding in detector operations and upgrades, presented for LIGO and Virgo observatories. Additionally, T. Dung Nguyen et al. (2023) fine-tuned the pretrained generative language model Llama-2 (H. Touvron et al. 2023) using 300,000 astronomy abstracts, demonstrating the model's ability to understand astronomical concepts through text completion. Further exploring the use of LLMs for astronomical information, V. Sotnikov & A. Chaikova (2023) present the way models like InstructGPT-3 (L. Ouyang et al. 2022) and Flan-T5-XXL (H. W. Chung et al. 2022) can automate the extraction of event IDs, event types, and astrophysical phenomena from The Astronomer's Telegram and Circulars.

In this study, we aim to explore the role that LLM-based NLP methods can play in modernizing the GCN. We aim to automate extracting relevant information about transient events from the Circulars archive. To achieve this, Section 2 focuses on a neural topic modeling pipeline for the GCN Circulars archive with the help of the library BERTopic (M. Grootendorst 2022). This pipeline automatically identifies common astrophysical themes in the Circulars archive and summarizes them with the help of an open-source LLM. We further enhance this pipeline by fine-tuning a BERT-based model to classify Circulars based on the type of reported observation. Section 3 describes the design and implementation of an open-source, end-to-end, zero-shot system designed for the automated extraction of GRB redshift information. The zero-shot approach enables the model to perform effectively without any prior examples. This system enables us to extract redshift values or ranges, GRB numbers, names of telescopes or observatories, and redshift types from the GCN Circulars archive. Additionally, we conducted a preliminary quantitative analysis of the data obtained. Our approach utilizes open-source models that require no training, allowing them to be easily adapted to extract various types of astrophysical information by simply fine-tuning the prompts. We have also implemented a simple retrieval-based system to filter out Circulars that lack the relevant information, which significantly improves the accuracy and reduces computational costs.

[12] https://gcn.nasa.gov/notices
[13] https://gcn.nasa.gov/Circulars
[14] https://www.wis-tns.org/
[15] https://minorplanetcenter.net/

## 2. Topic Modeling

### 2.1. Methods

#### 2.1.1. Neural Topic Modeling with BERTopic

In the field of NLP, topic modeling is a statistical text-mining technique used to uncover latent themes present within large text corpora. This method enables the identification of semantically meaningful clusters, known as topics, in extensive text datasets (H. Zhao et al. 2021). Over the years, various probabilistic topic modeling techniques based on Bayes' theorem, such as latent Dirichlet allocation (D. M. Blei et al. 2003) and Probabilistic Latent Semantic Analysis (T. Hofmann 2013), have been successfully employed. More recently, deep neural network architectures, particularly the transformer model (A. Vaswani et al. 2017), have been applied to create more semantically meaningful topic clusters (S. Sia et al. 2020; M. Grootendorst 2022). These algorithms utilize word embeddings, which are multidimensional vector representations of words. Pretrained LLMs, such as BERT (J. Devlin et al. 2018), are used to create these word vectors in such a way that semantically similar words are positioned close to each other within the vector space. Entire sentences and even documents can be encoded by averaging over all the word embeddings in that document (N. Reimers & I. Gurevych 2019). The document embeddings can then be clustered together using clustering algorithms like k-means or Hierarchical Density-Based Spatial Clustering of Applications with Noise (HDBSCAN; L. McInnes & J. Healy 2017) into groups of semantically related documents, or topics.

In this work, we employ the topic modeling library BERTopic (M. Grootendorst 2022) to build a neural topic model for the Circulars archive. A standard BERTopic pipeline involves first converting a collection of documents into numerical vectors, usually achieved with a Sentence Transformers model (N. Reimers & I. Gurevych 2019) based on the BERT architecture. These models encode semantic and contextual information from each document into high-dimensional document vectors, also known as embeddings. The high dimensionality of these vectors, however, makes it challenging to accurately cluster them using conventional clustering algorithms due to higher sparsity in the data, sometimes called the "curse of dimensionality" (K. Beyer et al. 1999; C. C. Aggarwal et al. 2001). Thus, a dimensionality reduction algorithm is applied to the document vectors. Common algorithms used for this purpose are principal component analysis (PCA; H. Abdi & L. J. Williams 2010) and Uniform Manifold Approximation (UMAP; L. McInnes et al. 2018). Finally, our reduced embeddings are grouped into coherent topic clusters using an appropriate clustering algorithm. The default algorithm used for this purpose is HDBSCAN, a robust clustering algorithm that can discover hierarchical clusters of arbitrary shapes and sizes within high-dimensional data by grouping neighboring data points with sufficiently high densities (R. J. G. B. Campello et al. 2013; L. McInnes & J. Healy 2017).

Once the topic clusters have been extracted from the dataset, these topics can be represented in multiple ways. By default, BERTopic characterizes topics through the most significant words present in each cluster. To determine these key terms, a variation of the Term Frequency-Inverse Document Frequency (TF-IDF), formula is employed (T. Joachims 1997). Called `c-TF-IDF`, where `c` stands for class-based, identifies the most informative keywords in each cluster by concatenating all documents within them and treating them as a single text (M. Grootendorst 2022). Additionally, we have the option of removing unimportant words being considered keywords by implementing a custom stopword list. This is a list of uninformative words that do not contribute to the interpretability of our extracted topics and should thus be removed from our representations. An example of common stopwords we chose to remove from our astrophysical topic representations includes articles like "a," "an," and "the," as well as author names, website URLs, and emails.

#### 2.1.2. Unsupervised GCN Topic Modeling Pipeline

The GCN platform provides access to its entire archive of Circulars as a downloadable archive on its website.[13] Circular data is available in both text and JavaScript Object Notation (`JSON`) formats. The advantage of using `JSON` format is that it organizes data into key-value pairs and can be automatically parsed by machines with minimal manual intervention. For our study, we downloaded the `JSON` tarball file and extracted the Circular bodies and subject headers from the "body" and "subject" `JSON` fields, respectively. We also extracted the date on which each Circular is published from the "createdOn" fields for later trend analysis. Circular data was collected for every GCN Circular published between 1998 March and 2025 May. For each Circular, we created a single text by concatenating the subject and body. This list of each circular text was then processed by removing any undefined characters or symbols to facilitate easier processing by our pipeline. Some preliminary word statistics were collected, and a list of the most common uninformative stopwords was manually curated. The list includes some commonly occurring words in Circulars, such as "acknowledgment" and "scheduled," as well as names of authors and organizations that do not provide relevant information on astrophysical topics of interest. Thus, we curated a custom set of stopwords of 1587 terms. Additionally, a general list of website URLs and numbers was also extracted from the circular texts using regular expressions and string pattern matching, respectively. Some emails were collected from the "email" fields of the Circular `JSON` files as well. These entities were added to our manually curated list, which includes a predefined list of stopwords from the Python `NLTK` library (S. Bird et al. 2009) and a punctuation list from Python's string module.

Next, each Circular text was then encoded into a vector representation with an embedding model from the Sentence Transformers library (N. Reimers & I. Gurevych 2019). We used the default `all-MiniLM-L6-v2` model, which is a small but highly effective embedding model shown to have comparable performance to much larger models for neural topic modeling (M. Grootendorst 2022). Other embedding models were also tested; however, we found that the quality of topics generated, evaluated through manual assessment, showed negligible change to justify the higher computational costs for larger embedding models.

Finally, our Circular texts, custom stopword list, and circular embeddings were fed into our BERTopic pipeline for dimensionality reduction, clustering, and keyword extraction.

```
Q: I have a topic that contains the following representative circulars among hundreds of others:
[DOCUMENTS]

The topic is described by the following keywords: '[KEYWORDS]'.

Based on the above information, can you give a short label of the topic?
A:
```

**Figure 1.** Prompt template used for topic summary generation. For every topic cluster [DOCUMENTS] gets replaced by three concatenated sample Circulars from each topic, while [KEYWORDS] gets replaced by the representative keywords extracted using `c-TF-IDF`. Prompt template modified from the default template provided in BERTopic documentation.

We used the default UMAP and HDBSCAN algorithms for dimensionality reduction and clustering, respectively. Preliminary tests with PCA for dimensionality reduction were performed; however, they resulted in a significant drop in topic quality. The default HDBSCAN algorithm was selected for clustering because it does not require the user to specify the number of clusters in advance. Additionally, it can identify outliers in the dataset that do not fit into any topic. This feature makes it a preferred choice for topic discovery over the GCN archive, where we expected to find various errata and miscellaneous Circulars that do not belong to any specific astrophysical topic. Topic keywords were extracted with `c-TF-IDF`, where our custom stopword list was used to filter out uninformative keywords.

To generate the embeddings for clustering, we employed UMAP for dimensionality reduction. The model was configured with `n_neighbors` as 30 to get balanced local and global structure, enabling the separation of broad astrophysical categories such as GW events, GRBs, neutrino detections, and multimessenger follow-ups (for scientifically balanced topic recommended range is 25–50). The embeddings were reduced to `n_components` as five dimensions to capture sufficient variance, and `min_dist` as 0.0 for dense packing of points, which enhances cluster distinction. Cosine distance (metric = "cosine") was used to preserve semantic similarity between textual embeddings, and the random seed was fixed (`random_state` = 42) to ensure reproducibility across runs.

Following dimensionality reduction, clusters were extracted using HDBSCAN, with parameters tuned for coarse, interpretable topic modeling. HDBSCAN offers two such parameters that can significantly affect topic modeling performance: `min_cluster_size` and `min_samples`. The `min_cluster_size` parameter determines the minimum size of the smallest topic cluster. Increasing this value results in the merging of smaller clusters and consequently reduces the overall number of topics. Ideally, we want to maintain a manageable number of topics for further analysis while avoiding the formation of small microclusters with only a few documents. Through preliminary analysis, we found that setting this hyperparameter to 800 consistently yielded between 10 and 20 topics from our dataset. Whereas the `min_samples` hyperparameter controls the degree to which certain data points are classified as noise. We prefer a lower value for this parameter to ensure that as many Circulars as possible are clustered together. After conducting initial tests, we decided to set this hyperparameter to 10, which minimized the number of Circulars declared as outliers.

### *2.1.3. Topic Summarization with `Mistral 7B Instruct`*

To enhance the interpretability of astrophysical topics, more informative topic representations are built using generative LLMs. By feeding in a few sample documents from each topic, the topic keywords obtained from `c-TF-IDF` and a carefully designed prompt template for an LLM, we guide the AI to provide brief, informative summaries of the underlying scientific descriptions. This method can help us build much more informative topic labels for each topic cluster in contrast to standard methods like `c-TF-IDF` alone. BERTopic also provides support for integrating LLMs in this manner into our pipeline. For topic summary generation, we selected the open-source `Mistral 7B Instruct` model (A. Q. Jiang et al. 2023). This 7.3 billion parameter pretrained generative model has been fine-tuned on instruction datasets from Hugging Face.[16] It has shown high performance on several text generation benchmarks, exhibiting excellent performance in comprehension and reasoning benchmarks. Given the hardware constraints of running our model on a single Google Colab GPU environment, we used an even smaller version of this model in which the weights of the model were "quantized" or reduced to 4 bits. This version of the model, called `mistral-7b-instruct-v0.2.Q4_K_M.gguf` (T. Jobbins 2023), was downloaded from the Hugging Face library[17] and run in Colab with the help of the open-source project `llama.cpp` (G. Gerganov 2023). The prompt template that we used for summary generation is presented in Figure 1.

### *2.1.4. Supervised Observation- and GW-based Classifications*

Our standard topic modeling pipeline allows us to automatically identify themes present in the Circulars archive. Next, we attempted to cluster our Circulars based on predefined astrophysical categories of our own choosing. More specifically, we categorized Circulars into five broad categories depending upon the type of observation being reported in them. Our five categories were:

1. *High-energy observations.* This includes X-ray observations as well as gamma-ray observations across the MeV, GeV, and TeV energy ranges. Telescopes in this category include Fermi Gamma-ray Burst Monitor (Fermi GBM; C. Meegan et al. 2009), Fermi Large Area Telescope (Fermi-LAT; W. B. Atwood et al. 2009), Swift Burst Alert Telescope (Swift-BAT; N. Gehrels et al. 2004), Swift X-ray Telescope (Swift/XRT), Nuclear Spectroscopic Telescope Array (F. A. Harrison et al. 2013),

[16] https://huggingface.co/

[17] https://huggingface.co/TheBloke/Mistral-7B-Instruct-v0.2-GGUF/tree/main

High Energy Stereoscopic System (J. Hinton 2004), High-Altitude Water Cherenkov Gamma-Ray Observatory (R. Springer 2016), and many more.

2. *Optical observations.* This includes Circulars with reported near-infrared (NIR) and ultraviolet (UV) detections, spectroscopic and photometric redshifts, as well as optical magnitudes and upper limits. Optical telescopes include Hubble Space Telescope (H. C. Ford et al. 1998), James Webb Space Telescope (J. P. Gardner et al. 2006), Zwicky Transient Facility (ZTF; E. C. Bellm et al. 2018), Nordic Optical Telescope (NOT; J. Vernin & C. Muñoz-Tuñón 1992), W. M. Keck Observatory (Keck; J. E. Nelson et al. 1985), and Very Large Telescope (VLT; J. Vernet et al. 2011), etc.
3. *Radio observations.* This includes detections of radio sources by observatories and telescopes such as the Very Large Array (VLA; P. J. Napier et al. 1983), Arcminute Microkelvin Imager (J. Zwart et al. 2008), Australia Telescope Compact Array (W. E. Wilson et al. 2011), Atacama Large Millimeter/submillimeter Array (ALMA; A. Wootten & A. R. Thompson 2009), and Giant Metrewave Radio Telescope (G. Swarup et al. 1991), etc., with flux densities often measured in Jansky and millijansky units.
4. *GW observations*. This includes data from laser interferometers such as LIGO, Virgo, and KAGRA (B. P. Abbott et al. 2018).
5. *Neutrino observations.* Observations reported by neutrino detectors such as IceCube (M. G. Aartsen et al. 2017) and the ANTARES neutrino telescope (M. Ageron et al. 2011) are classified here.

Categorizing Circulars based on the type of observation is challenging with conventional methods due to the lack of suitable keywords with which to sort them. For example, a search for the word "radio" in the GCN Circulars archive will often turn up many Circulars from the GRBAlpha CubeSat, even though they have nothing to do with radio observations, as they all end with the statement *"The ground segment is also supported by the radio amateur community and it takes advantage of the SatNOGS network for increased data downlink volume,..."* referring to radio communications for their CubeSat. Transformer-based language models can be used here to take into account contextual information and thus improve the accuracy of simple keyword-based sorting methods. We demonstrate this capability by modifying our topic modeling pipeline to perform a type of neural search over the GCN archive with the help of BERTopic's zero-shot topic modeling technique.[18] A list of five candidate labels—"High Energy Observations," "Optical Observations," "Radio Observations," "Neutrinos," and "Gravitational Wave" was passed into our pipeline and embedded with the help of the `all-MiniLM-L6-v2` embedding model. The embedding vectors of these labels were then compared to the embeddings of every Circular in our database using the cosine similarity metric. Each Circular was thus assigned to the label with the highest similarity score. Similarly, we conducted a comprehensive study on the classification of a specific type of GCN Circulars, focusing on classifying GW events, their electromagnetic (EM) or neutrinos counterparts, and non-GW-related astrophysical events. Such classification would be useful for the community to filter the heterogeneous stream of Circulars containing diverse events, thus facilitating rapid follow-up of targeted searches for specific classes of astrophysical phenomena. In this study, we focus specifically on GW alerts, using the candidate labels "Gravitational Wave," "Gravitational-Wave Counterpart," and "Non-Gravitational Wave" to classify these Circulars. To improve the accuracy of these methods on observation-based classification and GW-based classification, we opted to fine-tune our embedding model on a manually labeled dataset.

#### 2.1.5. Contrastive Fine-tuning

The performance of the observation-based clustering pipeline relies heavily on the ability of the `all-MiniLM-L6-v2` embedding model, a Sentence Transformer model beneficial for tasks like clustering or identifying sentence similarity. It is utilized to group Circulars that report similar types of observations, while keeping the dissimilar observation types away from each other in the vector space. However, this task is challenging as the model has not been explicitly trained on astrophysical data before. For instance, it may find difficulty in differentiating between a high-energy neutrino detection and a high-energy gamma-ray observation, as they both contain the word "high energy" in them. To address this issue and improve our pipeline's performance in transient alert classification, we fine-tuned our embedding model using a technique called supervised contrastive learning (B. Gunel et al. 2020). This approach employs a similarity-based loss function and utilizes a labeled dataset, allowing a pretrained language model to learn how to embed texts belonging to the same class closer to one another in the vector space, while distancing examples from different classes. This method has been shown to significantly improve the robustness of embedding models when applied to other domains (B. Gunel et al. 2020; X. Ma et al. 2021).

*Observation-based clustering.* We fine-tuned the embedding model on a manually labeled dataset of 200 Circulars to classify the astrophysical observations. Each Circular was labeled based on the type of observation reported: optical (e.g., telescopes such as ZTF, Pan-STARRS, etc.), radio (e.g., observations from ALMA or VLA), high energy (e.g., gamma-ray detections by Fermi-LAT or Swift), GW (e.g., LIGO/Virgo/KAGRA events), and neutrino observations (e.g., IceCube detections). It was implemented using the Sentence Transformers library that specializes in optimizing text embeddings through contrastive fine-tuning, which offers support for various types of loss functions. Here, we used the distance-based contrastive loss function (R. Hadsell et al. 2006), which optimizes embeddings by minimizing the squared Euclidean distance between samples of the same class while maximizing it for dissimilar pairs. The fine-tuned `all-MiniLM-L6-v2` models were then used to create new embeddings for Circulars. These embeddings were fed into our observation-based clustering pipeline, which groups Circulars by similarity along with the observational labels. We found that our classification accuracy improved significantly upon using this fine-tuning approach.

*GW classification.* We fine-tuned the `all-MiniLM-L6-v2` embedding model using a manually labeled dataset consisting of a representative subset of 300 Circulars. This dataset includes a balanced set of 100 Circulars reported by LIGO, Virgo, and KAGRA; 100 follow-up observations of GW counterparts reported in GCN; and 100 Circulars unrelated to GW detections. The fine-tuning process was again implemented using the Sentence Transformers library and contrastive loss function. This approach significantly improved the model's ability to distinguish between GW-related and non-GW observations, achieving high classification accuracy.

[18] https://maartengr.github.io/BERTopic/getting_started/zeroshot/zeroshot.html

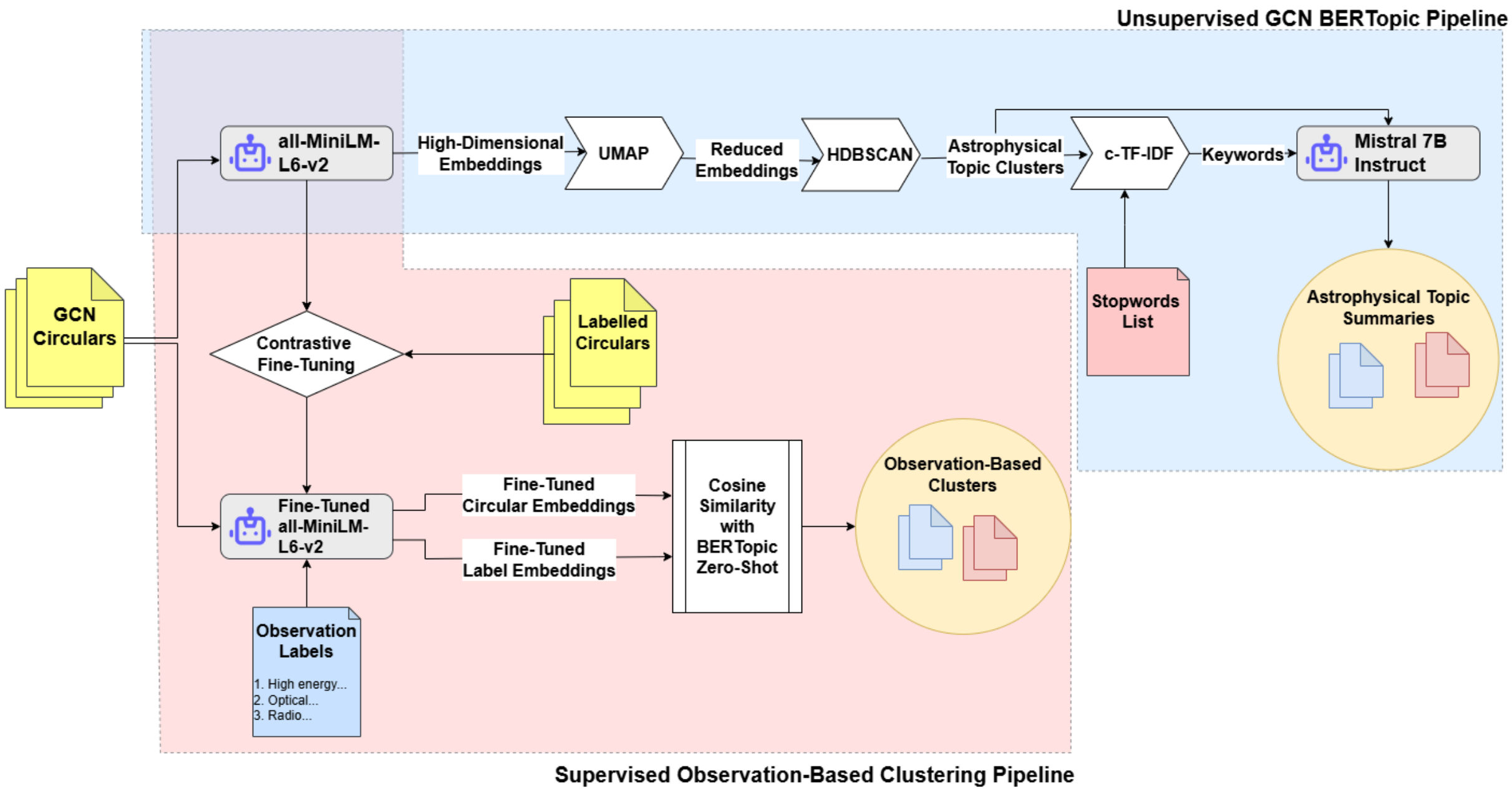

**Figure 2.** Visualization of the GCN topic modeling pipelines, built with the help of the BERTopic library. The blue pipeline depicts the steps for the unsupervised discovery of astrophysical topics. Circulars are embedded with `all-MiniLM-L6-v2`, after which the vectors are reduced and clustered. Keywords are extracted after stopwords removal, and the discovered topics are summarized with `Mistral 7B Instruct`. The red pipeline describes the supervised process for generating observation-based clusters. The `all-MiniLM-L6-v2` model is fine-tuned on a labeled dataset of Circulars. The Circulars and a list of observation labels are then embedded using this model. Finally, both sets of embeddings are compared with each other using the cosine similarity metric to find the closest observation label for each Circular in the embedding space.

Figure 2 provides the flowchart for both unsupervised and supervised learning in the blue-shaded region and red-shaded region, respectively, for our topic modeling pipeline.

### 2.2. Results

The results of applying LLMs to GCN Circulars are systematically stored and presented in the NASA-GCN GitHub repository[19] at `nasa-gcn`, serving as an open resource for reproducibility and access to the products generated in this work. The repository contains all data, analysis pipelines, visualizations, and extended tabular results. The repository is organized into three primary folders: data, scripts, and extended tables. The data directory contains archived Circulars in JSON format (up to 2025 May) and a custom stopword list used for preprocessing. The scripts directory provides Google Colab notebooks for topic modeling. The extended tables directory contains tabular outputs, including topic modeling cluster tables for GW-focused topics, observational clusters, and an unsupervised topics summary.

#### 2.2.1. GCN Topic Clusters and Summaries

The pipeline consists of three stages: embedding generation, topic modeling, and summarization. Embedding all Circulars with `all-MiniLM-L6-v2` took approximately a minute, demonstrating its suitability for large-scale semantic representation studies. Topic modeling by itself, which includes the dimensionality reduction, clustering, and keyword extraction phases, was completed in a few minutes. Finally, topic summaries were generated with `Mistral 7B Instruct` in about a minute as well.

Using our topic modeling pipeline with our chosen hyperparameters, we get 24 unique topics from the GCN Circulars archive, not counting the documents labeled as outliers. After clustering our Circulars, we visualize our topic clusters in 2D after reduction of our document embeddings with `t-SNE`, as seen in Figure 3. Note that topics here are represented using their keywords extracted with `c-TF-IDF`. Finally, 13 topic summaries generated by `Mistral`, along with their Circular counts, can be viewed in Table 1.

#### 2.2.2. Fine-tuning Setup, Observation-based Classification, and Evaluation

To generate observation-based clusters from Circulars, we employed a contrastive learning approach to fine-tune the `all-MiniLM-L6-v2` sentence embedding model. As mentioned previously, the training dataset consisted of 200 manually categorized Circulars, evenly distributed across five observation types: high-energy, optical, radio, neutrino, and GW detections, with 40 Circulars for each category. The dataset was divided into training and test sets using an 80–20 split, resulting in 160 training samples and 40 test samples. In the contrastive fine-tuning setup, each Circular in our training set was paired with every other Circular, leading to 25,600 text pairs. We also include category labels, "Optical Observations," "High Energy Observations," "Radio Observations," "Neutrinos," and "Gravitational Wave" for each observation type in the training set. This approach allowed the model to learn to represent both the text of the Circulars and their corresponding labels together within the same latent vector space. We thus had a total of $(32 \text{ High Energy} + 32 \text{ Optical} + 32 \text{ Radio} + 32 \text{ Neutrino} + 32 \text{ GW} + 5 \text{ Labels})^2 = 27{,}225$ final text pairs. Each pair was assigned a binary label: a label of 1 was assigned to pairs where both Circulars (or Circular and label) belonged to the same observation type (positive pair), while a label of 0 was given to pairs that belonged to different observation types (negative pair). These labeled pairs were used to optimize the embedding model using a contrastive loss function, implemented via the Sentence Transformers library. Training was conducted over 1 epoch, which we found produced the best results; more epochs did not lead to improvement. The time required for fine-tuning was approximately half an hour on a T4 GPU (Google Colab) per epoch.

[19] https://github.com/nasa-gcn/circulars-nlp-paper

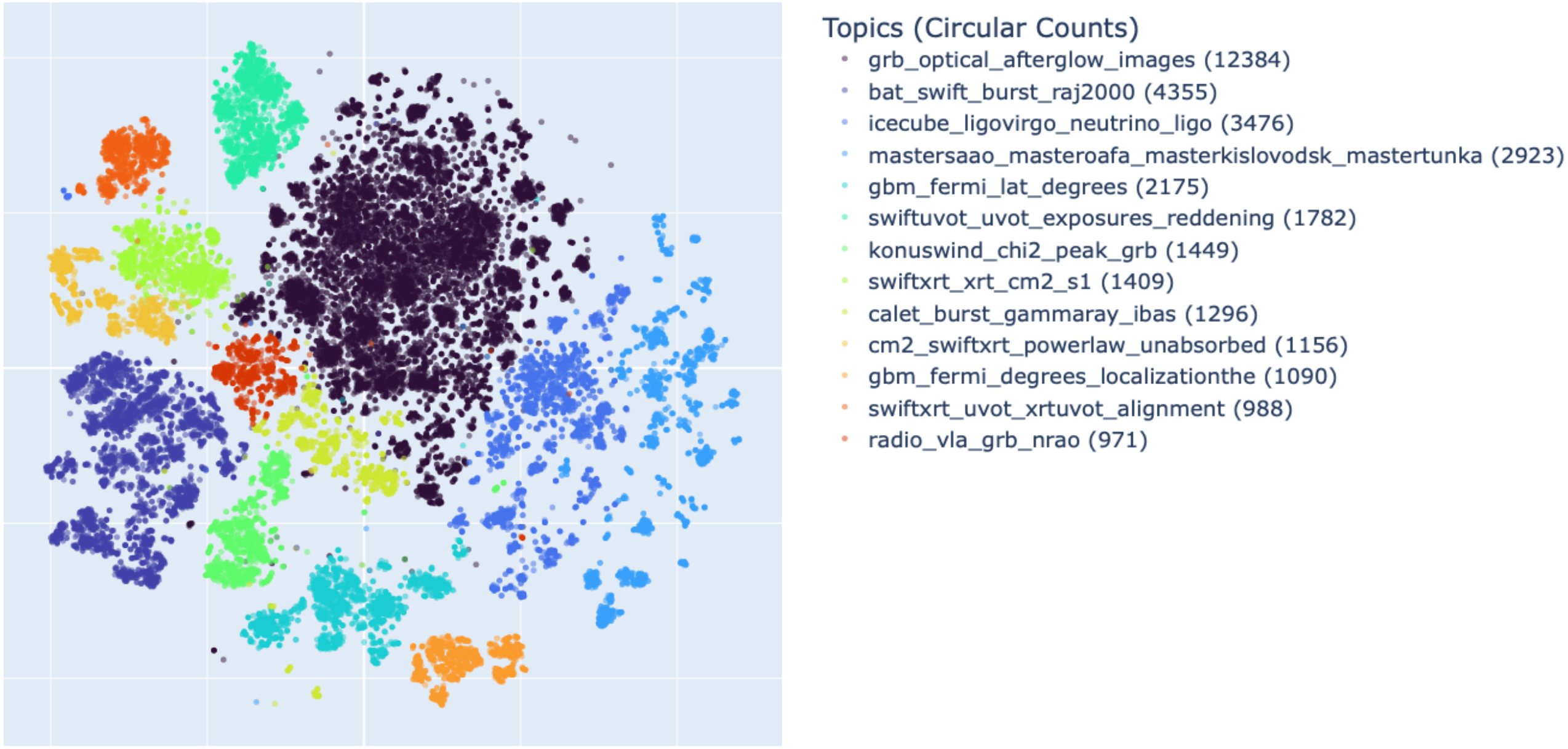

**Figure 3.** GCN topic clusters after reduction with `t-SNE` with unsupervised pipeline: topics on the right are represented with their four most important keywords as extracted with `c-TF-IDF`. Next to the topic labels in parentheses are the number of Circulars each topic contains. Note that outliers were excluded from the representation. The topics are also available in Table 1.

**Table 1**
Topic Summaries Generated with `Mistral 7B Instruct`

| Topic Summary | Circular Count |
|---|---|
| GRB Observations and Afterglow Detections with Various Telescopes. | 12,384 |
| Swift Detection and Analysis of Gamma-Ray Bursts: BAT Trigger Data and XRT Observations. | 4355 |
| LIGO/Virgo/KAGRA Detection of Compact Binary Merger Candidates and Their Properties. Also mentions the IceCube Collaboration and neutrino detection. | 3476 |
| MASTER-Net Observations of GRBs by Multiple Robotic Telescopes: MASTER-SAAO, MASTER-OAFA, and Others. | 2923 |
| Fermi GBM Detections: GRB Light Curves, Spectral Analysis, and Event Fluence. | 2175 |
| Swift/UVOT Upper Limits for GRBs: White Exposures, $3\sigma$, Uncorrected for Galactic Extinction. | 1782 |
| Konus-Wind Observations of Long-Duration GRBs: Light Curves, Spectra, and Fitting Results. | 1449 |
| Swift-XRT Team Reports X-Ray Source Detections and Analysis for GRBs. | 1409 |
| CALET Gamma-Ray Burst Monitor Detections: GRB 200805A, GRB 170703A, GRB 180720B, and GRB 170702A. | 1296 |
| Swift-XRT Analysis of GRB Light Curves and Spectra: Power-law Decays and Column Densities. | 1156 |
| Fermi GBM GRB Localization Reports: R.A., Decl., Statistical Uncertainty, Skymap, HEALPix, and Angle from Fermi-LAT Boresight. | 1090 |
| Swift-XRT Team Reports Astrometrically Corrected X-Ray Positions for GRBs with Uncertainties and Improvement Process. | 988 |
| Radio Observations of GRBs with the VLA at 8.46 GHz Frequency. | 971 |

Both the original (pretrained) and fine-tuned embedding models were applied to cluster the Circulars archive using BERTopic's Zero-Shot pipeline, which checks for the cosine similarity between an observation label and a Circular and assigns labels accordingly. To ensure comprehensive labeling, we set the cosine threshold in the pipeline to 0.1, ensuring that every Circular was assigned to at least one category. Model performance was quantified by computing classification accuracies on the training and test sets for the embedding model, both before and after training. It is estimated as the fraction of correctly labeled Circulars in the training and test sets.

The results demonstrate that contrastive fine-tuning significantly enhanced the classification accuracy of our pipeline. Specifically, the accuracy on the training set increased from 66.9% (pretrained) to 100% (fine-tuned), while the accuracy on the test set increased from 65% (pretrained) to 90% (fine-tuned). Test accuracy declines to 87.5% and 85.0% at epochs 2 and 3, respectively, indicating overfitting with continued training. Thus, we utilized this fine-tuned model to classify

(a)

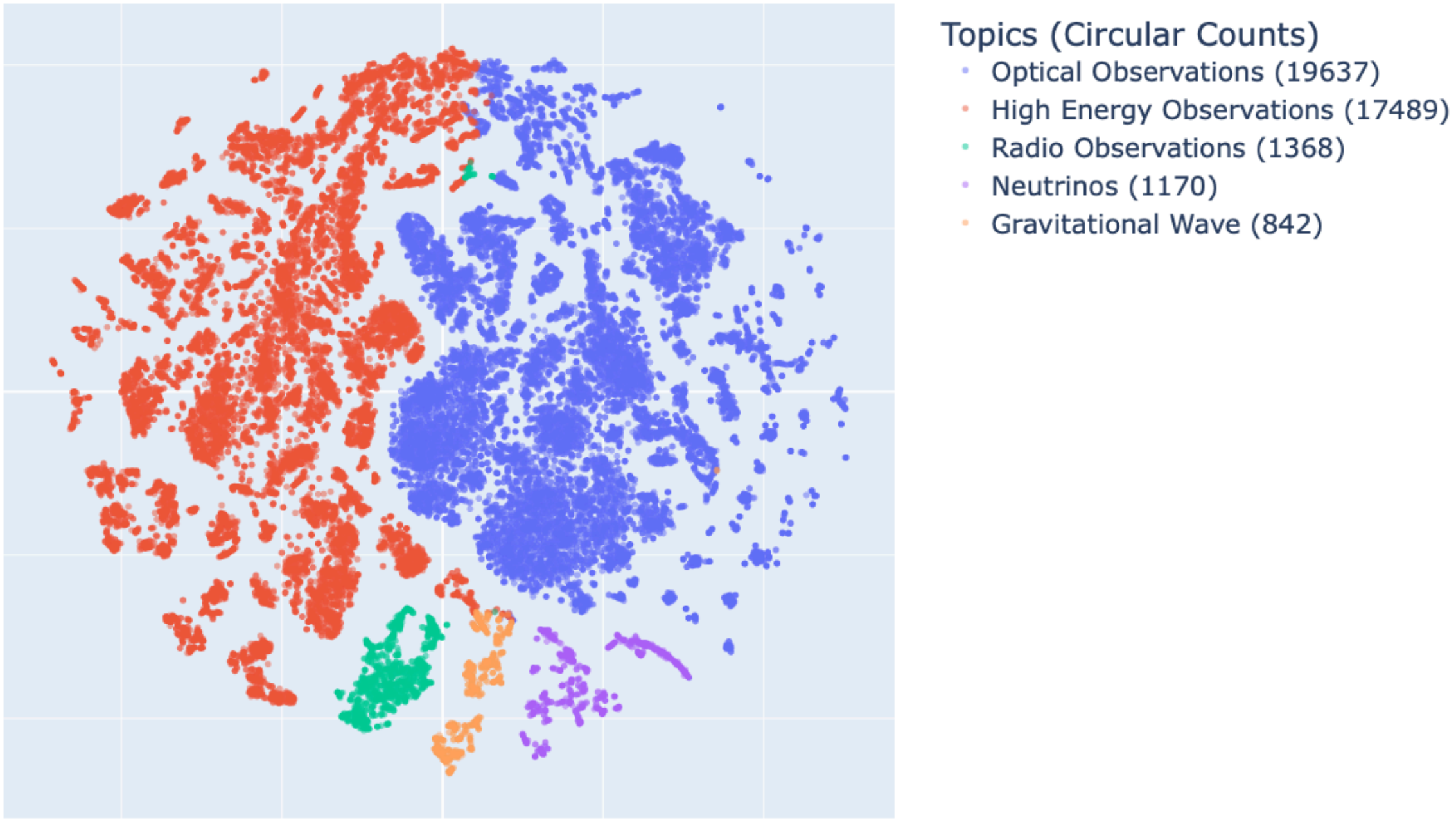

(b)

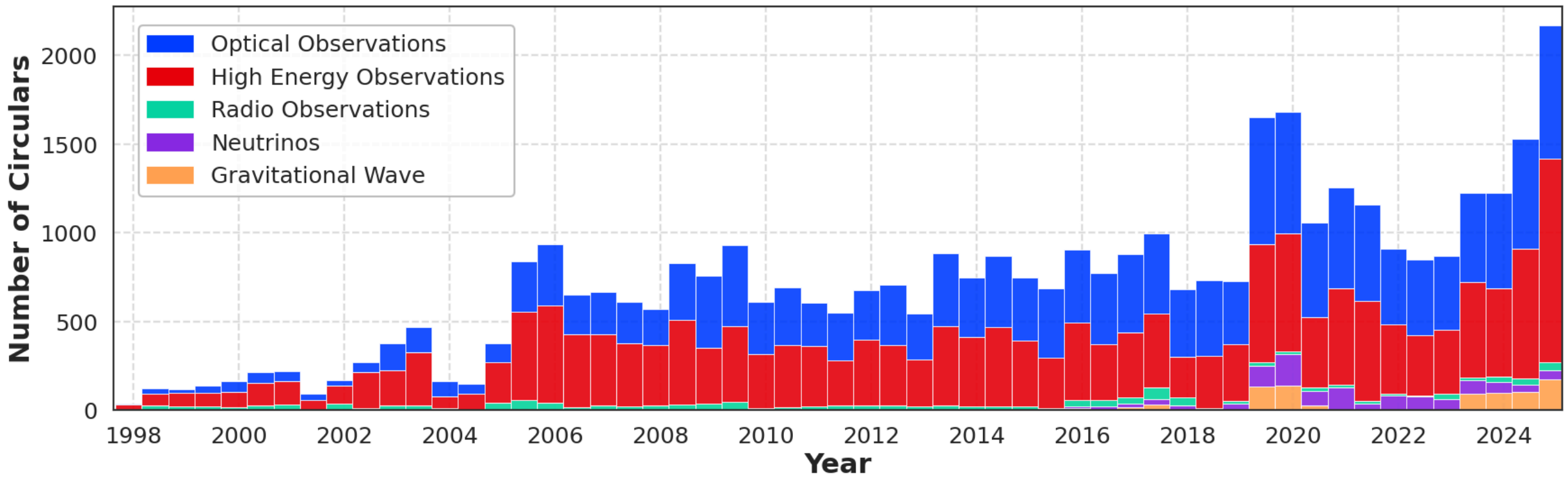

**Figure 4.** (a) `t-SNE` representation of GCN observation-based clusters. The embedding vectors of each circular are classified using cosine similarity between embedded labels and documents. Contrastive fine-tuning helps our model more accurately classify the circular embeddings. (b) Trends over time of the observation-based circular clusters generated through supervised contrastive fine-tuning of our embedding model (`all-MiniLM-L6-v2`).

the Circulars into observation-based clusters, as can be seen in the `t-SNE` plot in Figure 4(a), which presents well-separated observation-based clusters. Additionally, the temporal trends among these clusters were analyzed by extracting the submission dates from each Circular and plotting the observation frequencies over time, as shown in Figure 4(b).

Figure 4(a) shows the clustering of GCN Circulars by wavelength and messenger type, grouped into five categories: high energy, optical, radio, GW, and neutrinos. The largest number of reports in GCN come for the high-energy and optical observatories, highlighted in red and blue. The radio observations are shown in green, while GW detections, shown in orange, account for only a small fraction (beginning in 2015). Neutrino observations are represented in purple. To analyze the behavior of these machine learning (ML)-derived clusters, we performed a trend analysis for each of these categories, as shown in Figure 4(b). We clearly see that the GW cluster started in 2015, after which follow-up observations increased and led to the rise in the other clusters as well. These results demonstrate the effectiveness of LLM-based clustering in capturing the evolution of wavelength- and messenger-specific activity within the Circulars.

### *2.2.3. GW and GW Counterpart Classification*

We investigated the classification of GW Circulars using the fine-tuned `all-MiniLM-L6-v2` language model. The model was fine-tuned on a balanced dataset of 300 Circulars (100 per label), categorized into three classes: GW Circulars, GW Counterpart Circulars, and all other Circulars labeled as Not Gravitational Wave. The fine-tuning process utilized a contrastive learning framework, similar to previous setups, with an 80–20 train–test split. We also include “category labels,” specifically, “Gravitational Wave,” “Gravitational-Wave Counterpart,” and “Non-Gravitational Wave,” in the training set. In the contrastive learning framework, the total number of possible pairs was thus (80 GW + 80 GW counterparts + 80 non-GW + 3 labels)$^2$, which equals a total of 59,049 text pairs. Due to the large dataset, the training was conducted for one epoch, completed in approximately 1 hr on Google Colab.

The classification accuracies for both the training and test sets in our zero-shot BERTopic pipeline, which assigns Circulars to the three predefined categories, were evaluated. The base `all-MiniLM-L6-v2` model shows limited performance, with accuracies of 23.3% for training and 25.0% for testing. After fine-tuning for a single epoch, the model achieves substantial improvement, reaching 100.0% accuracy on the training set and 98.3% accuracy on the test set. The results indicate that despite the model being trained for just one epoch, it successfully distinguished between the three categories. Clusters of all Circulars based on this fine-tuned model are shown in the `t-SNE` plot in Figure 5(a). A clear distinction can be visually observed between Circulars belonging to the GW, GW counterpart, and non-GW categories in our plot. As a test, Circulars that reported observations on GW170817 were also labeled and plotted, shown in yellow. Since GW170817 was extensively followed up by the ground and space-based observatories, the vast majority of these Circulars consist of the GW counterparts, hence overlap with the corresponding cluster. 226 of these 227 Circulars were correctly classified as belonging to either the GW or the GW counterpart cluster. One Circular, number 29041 (A. Hajela et al. 2020), initially identified as non-GW, is actually a GW counterpart. The evolution of all three clusters over time is illustrated in Figure 5(b). Thus, fine-tuning `all-MiniLM-L6-v2` with contrastive learning and label-augmented training data leads to highly accurate and generalizable clustering of Circulars.

Figure 5(a) presents the clustering of GCN Circulars focused on GW events. Only GW detection reports by the LIGO/Virgo/KAGRA Collaboration are shown in green, while clusters corresponding to GW EM counterparts are shown in red, basically follow-up Circulars of GW detection. All other remaining Circulars not linked to GW events, which include GRB, TDE, blazar, and SGR events, are shown in blue. To highlight the behavior captured by the LLM, we additionally just overplot (not ML clustering) Circulars associated with the binary neutron star merger GW170817/GRB 170817A event in yellow color. This particular event, with GW detection with a confirmed EM counterpart, demonstrates a dense concentration of follow-up reports overlapping the red GW counterpart cluster by LLM, as intensively followed up by multiwavelength observatories worldwide. To further understand the ML clusters, we examined the temporal evolution of these clusters as shown in Figure 5(b). The results confirm that both GW and GW counterpart clusters emerge only after 2015, coinciding with the start of the GW era with the first GW detection from a binary black hole merger, GW150914 (B. P. Abbott et al. 2016), reported by the LIGO and Virgo Scientific Collaboration. In contrast, Circulars unrelated to GW events, as shown in blue, have been reported consistently since the start of GCN. These findings illustrate the effectiveness of LLM-based clustering in tracing the evolution of multi-messenger follow-up activity.

## 3. Information Extraction

### 3.1. Methods

#### 3.1.1. Zero-shot Information Extraction

Information extraction is a technique used to extract structured data from unstructured texts. These unstructured sources may include journal articles, scientific papers, and even Circulars, which are written by humans for humans, and thus follow a flexible format that makes them difficult for machines to parse. Information extraction techniques are beneficial for automatically processing large volumes of text and extracting structured information from them, which can be further analyzed and interpreted. Traditionally, this used to be done with manually designed rule-based systems to extract named entities, events, or relationships (J. R. Hobbs & E. Riloff, 2010), and required extensive customization. Recently, advances in ML and NLP have transformed this field. Instead of handcrafted rules, modern techniques use learning-based models that automatically recognize patterns in text and extract the required entities. Subsequently, LLMs have advanced information extraction by demonstrating strong performance in converting scientific texts into structured data (M. Agrawal et al. 2022; A. Dunn et al. 2022; V. Sotnikov & A. Chaikova 2023; Z. Zheng et al. 2023). A key advantage of using LLMs is their ability to learn domain-specific tasks that they were not explicitly trained on, an ability commonly referred to as zero-shot learning (T. B. Brown et al. 2020; J. Wei et al. 2021). Zero-shot learning allows LLMs to process vast quantities of domain-specific literature, which in our case are Circulars, and extract all relevant information within them without the need for curating custom datasets for each specific task. Furthermore, it has been found that the effectiveness of these models on such tasks can be improved through prompt engineering (J. Wei et al. 2021), a systematic design of input (or “prompt”) for NLP models to optimize their outputs. This process involves instructions, examples, or queries that guide the model to generate contextually relevant responses. For our study, we implemented an open-source LLM and employed carefully tuned prompts to develop a zero-shot, end-to-end system to extract the astrophysical information from Circulars. We tested our system to extract the key astrophysical data: (1) redshift value/range reported for a GRB, (2) GRB event name, (3) telescope/observatory that observed the event, and (4) redshift measurement method; spectroscopy or photometry. Our approach requires no training of the model and thus can be readily applied to extract a wide range of astronomical data with simple modifications. Thus, by combining LLMs, zero-shot learning, and prompt engineering, we created a flexible and efficient system for extracting structured astrophysical data from unstructured Circulars.

#### 3.1.2. Model Selection

For information extraction, we employ the 4 bit quantized version of the `Mistral 7B Instruct` model, which is implemented with `llama.cpp`. At the time of its release, this

(a)

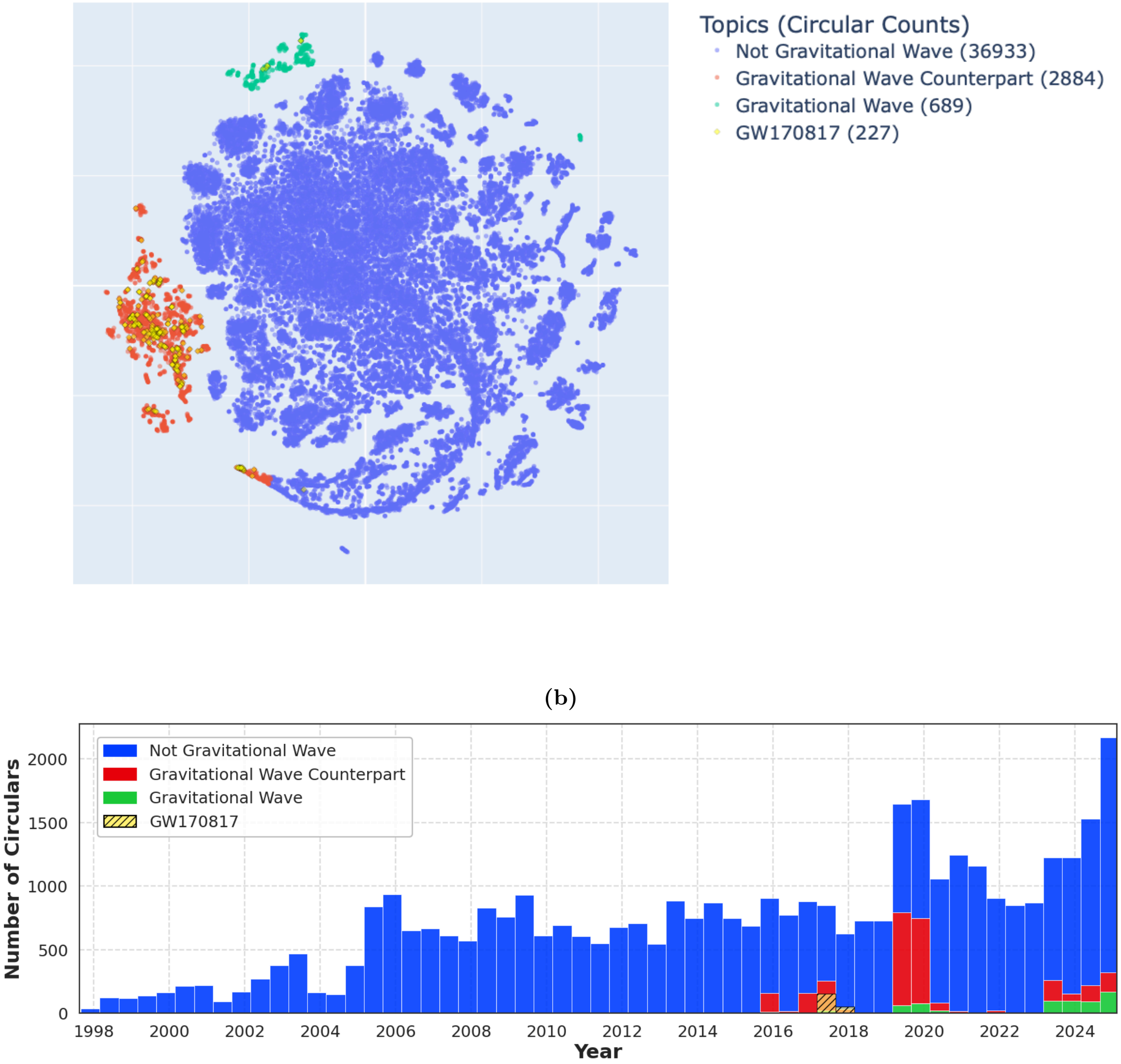

**Figure 5.** (a) `t-SNE` representation of "Gravitational Wave," "Gravitational-Wave Counterpart," and "Not Gravitational-Wave" clusters computed using contrastive fine-tuning of the `all-MiniLM-L6-v2` model. Green points represent machine-learned GW Circulars, red represent GW Counterpart Circulars, and blue represent other Circulars. Yellow diamond points mark Circulars associated with GW170817, as a representative example of cross-cluster overlap. (b) Topic trends over time for the same three clusters generated using supervised fine-tuning of `all-MiniLM-L2-v6. GW170817` is shown as a yellow hatched histogram.

model demonstrated remarkably high accuracy on several text generation benchmarks despite its relatively compact architecture (A. Q. Jiang et al. 2023). The quantization reduces its computational footprint even further, enabling efficient execution on a single instance of Google Colab's GPU. Preliminary evaluations on sample Circulars have shown promising results, particularly in extracting redshift information.

### *3.1.3. Prompt Engineering and Output Parsing with `LangChain`*

The output generated by an LLM is typically in the form of natural language. However, to perform data analysis for extracting the redshift values, the data must be presented in a more structured format. This is achieved by using fine-tuned prompts and appropriate output parsing algorithms. The `LangChain` library (H. Chase 2022) enables us to create a single prompt template that can be reused for all our Circulars. Additionally, it provides access to built-in modules that assist with structured output parsing. With some custom refinement, we developed a standardized template that can be applied to all Circulars to extract redshift information, as seen in Figure 6. Using this template, we process the Circulars with our quantized `Mistral 7B Instruct` model, which outputs the redshift information in `JSON` format.

Even with careful prompt engineering, the LLM may still occasionally produce results that deviate from our specified

```
GRB 151027A: Keck/HIRES redshiftD. A. Perley (DARK/NBI), L. Hillenbrand (Caltech), and J. X. Prochaska
(UCO/Lick) report:

We obtained a 900s spectrum of the optical afterglow of GRB 151027A
(Maselli et al., GCN 18478) with Keck/HIRES starting at UT 06:54 on
2015-10-27.  In the preliminary reduction, we identify a series of
strong absorption lines including the MgII doublet at z=0.81. Associated
MgI, FeII, and (likely) FeII* are also present, indicating that this is
the redshift of the GRB.  Analysis is ongoing.

For the given circular, can you extract the following information:

Redshift Value: Extract the redshift value observed for the GRB, if any. If not, answer No Redshift.If multiple measurements are present, only pick the most
likely one.
Telescope: What was the name of the telescope used for observation? Event: Extract the GRB Number of the event. Redshift Type: How was the redshift
determined? Answer Photometric if it was photometric. Answer Spectroscopic if it was spectroscopic or had absorption or emission features or lines. If
there's no redshift value, answer No Redshift. If there's no information on the type, answer No Information.

The output should be a markdown code snippet formatted in the following schema, including the leading and trailing "```json" and "```":

```json
{
        "Redshift Value": string  // Extract the redshift value observed for the GRB, if any. If not, answer No Redshift. If multiple measurements are
present, only pick the most likely one.
        "Telescope": string  // What was the name of the telescope used for observation? If not provided, answer No Information.
        "Event": string  // Extract the GRB Number of the event. If not provided, answer No Information.
        "Redshift Type": string  // How was the redshift determined? Answer Photometric if it was photometric. Answer Spectroscopic if it was spectroscopic
or had absorption or emission features or lines. If there's no redshift value, answer No Redshift. If there's no information on the redshift type, answer No
Information.
}
```

A:
```

```
```json
{
        "Redshift Value": "0.81",
        "Telescope": "Keck/HIRES",
        "Event": "GRB 151027A",
        "Redshift Type": "Spectroscopic"
}
```
```

**Figure 6.** Sample prompt used for extracting redshift information from a GCN Circular, in the top is the prompt/input, and at the bottom is the output from the 4-bit `Mistral 7B Instruct` model.

formatting rules, leading to parsing errors. To resolve this issue and ensure the seamless parsing of our extracted data, we postprocessed the LLM output with the help of Python regular expressions (regex) to fix any incorrect `JSON` strings. Regular expressions are sequences of characters used to match certain patterns of strings. For example, the regular expression "^GRB[0-9]+" or "^GRB [0-9]+" will match any string that begins with "GRB" followed by one or more digits (e.g., "GRB 1234," "GRB 9"). We analyzed the common error patterns in the LLM's output and applied regular expressions to match and adjust them to get properly structured `JSON`. Our postprocessing rules can be summarized as follows:

1. First, we removed all text outside of triple back-ticks "'...'". We stripped any comments, lines starting with "//".
2. Any incorrectly escaped backslashes "\" were fixed. Outputs indicating no redshift values, such as "N/A," "None," and "null," were standardized to "No Redshift."
3. For `JSON` syntax compliance, any keys or values not in double quotes "" were fixed. Missing commas were added at the end of every `JSON` key-value pair. Dangling commas at the end of the last `JSON` key-value pair were then removed.
4. Arrays in the model output were converted to single values for easier parsing. For example, ["value 1," "value 2"] would be converted to "value 1, value 2." Similarly, lists in the model output were also converted to single values. For example, "value 1," "value 2" would be converted to "value 1, value 2," as `JSON` requires single values per key.

We iteratively refined these postprocessing steps until the vast majority of our selected Circulars were successfully processed through our pipeline.

### *3.1.4. Testing and Evaluation with the Swift GRB Table*

The Neil Gehrels Swift Observatory (hereafter, Swift) maintains a manually curated archive that contains information on GRBs that are observed by Swift.[20] This table contains information from the Circulars. We use the redshift values and related GCN numbers from the archive as test data to evaluate the efficacy of our information extraction system. Data from the table was scraped, and only rows containing redshift values or ranges in the "Redshift" column were extracted. The reference Circulars for each redshift measurement were obtained from the "References" column of each row, which provides the GCN Circular numbers associated with the redshift detections. GRB numbers were also extracted from the archive's "GRB" column. Python regular expressions and string pattern matching were used to retrieve the redshift values, telescope or observatory names, and the types of redshift measurement from the "Redshift" column. The Swift GRB table generally reports redshifts with terms such as "absorption," "emission," or "photometric," indicating the measurement method. For analysis, we categorized both "absorption" and "emission" measurements as "Spectroscopic" redshifts. If no description was available, we parsed as "No Information" for that redshift measurement type.

In our evaluation table, rows containing multiple redshift measurements for a single GRB event were counted as separate entries, one for each measurement. In total, we extracted 540 redshift measurements for our evaluation table. Additionally, we included 105 Circulars that did not report any redshifts to evaluate our pipeline's performance on negative examples. Thus, in total, our evaluation table consisted of 645 data points. Circular numbers from this table were used to select the circular bodies for testing our LLM. As in our topic modeling pipeline before, the circular bodies were extracted from the Circulars archive after concatenating the subject and body fields in `JSON` format for each circular. Finally, we compared the redshift information extracted by our LLM pipeline from these Circulars with the information reported in the Swift table to evaluate the accuracy of the system.

[20] https://Swift.gsfc.nasa.gov/archive/grb_table/

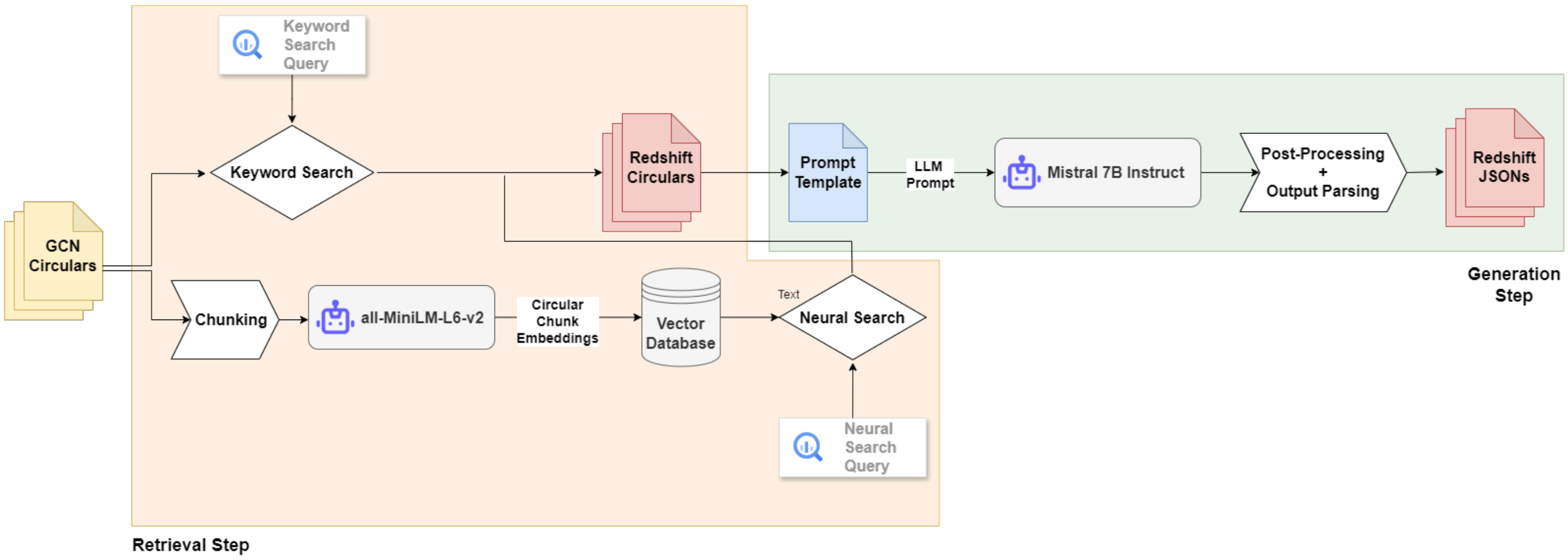

**Figure 7.** Visual representation of the redshift information extraction pipeline. Keyword search and neural search are combined to retrieve relevant redshift Circulars. For the generation step, retrieved redshift Circulars are combined with the tuned `LangChain` prompt template and fed into `Mistral 7B Instruct`. Finally, the generated output is processed to remove any formatting errors and parsed into `JSON` key-value pairs, with the fields containing the extracted redshift information from each Circular.

#### 3.1.5. Addressing LLM Hallucination in Redshift Information Extraction

Hallucination refers to instances where a language model generates unsupported, nonfactual, or nonsensical information in response to a prompt (J. Maynez et al. 2020; Z. Ji et al. 2022). This problem became particularly evident when we prompted our model with negative examples that lacked redshift values, but still contained ambiguous signals. Terms like "redshift" and the letter "$z$," which are commonly associated with redshifts, seemed to confuse the model. The cause for these errors may partly stem from the limitations of our smaller model. To mitigate this problem, we only prompt our model with Circulars that we are certain should include redshift observations. Thus, we retrieved only relevant Circulars from our database to feed into our pipeline. This method of enhancing LLM accuracy and reducing hallucination using text retrieval methods is commonly referred to as retrieval augmented generation or RAG (P. Lewis et al. 2020; Y. Gao et al. 2023).

In our implementation, we adopted a two-step approach to retrieve relevant documents. In the first step, we performed a straightforward keyword search within the subject field of all Circulars, looking for mentions of "redshift," "spectr," or "photo-$z$," along with the term "GRB." This method effectively retrieved the majority of GRB redshift-related Circulars. However, some Circulars in our database reported GRB redshifts without explicitly mentioning them in their subject field. To capture these, we next implemented a neural search mechanism with the assistance of `LangChain`. First, we split our remaining Circulars into smaller chunks to be more easily encoded by an embedding model. We then embedded these chunks using the small but efficient `all-MiniLM-L6-v2` Sentence Transformer. The resulting embedding vectors were indexed and stored in a locally hosted vector database designed for storing and querying over embeddings. Here, we used the Facebook AI Similarity Search library (M. Douze et al. 2025). Finally, we searched over this new database with the following query: "Spectroscopic Redshift $z$ Value with Absorption or Emission Lines." This query was first encoded using the same embedding model, after which it was compared to all Circular embeddings using a cosine similarity metric. Circulars that exceeded a certain similarity threshold, in our case 0.3, with the query were appended to our previously retrieved set. This method allowed us to gather the majority of relevant Circulars with reported redshift values, which we subsequently used for inference with our LLM. Figure 7 gives an overview of our redshift information extraction pipeline.

### 3.2. Results

#### 3.2.1. Evaluation of Redshift Information Extraction

The evaluation of the zero-shot information extraction pipeline was conducted in Google Colab using an L4 GPU instance. The total inference time for processing all 645 Circulars in our evaluation dataset was approximately 1.5 hr. Table 2 shows 10 randomly selected entries from our LLM-generated redshift information table, alongside the actual reported information from our evaluation dataset. 644 of the LLM output strings were successfully parsed into JSON objects without errors using structured output parsing and regex methods, and were thus added to our new evaluation dataset.

To assess the accuracy of our LLM-generated redshift information, we manually evaluated all 644 data points. We divided our dataset into two subsets: one subset consisted of positive examples, which were Circulars taken from the Swift GRB table that we knew contained redshift information. The second subset was comprised of the negative examples, which had no mention of reported redshifts. Thus, in total, we had 539 Circulars in the positive examples and 105 Circulars in our negative examples subset. In Table 3, we present the total

**Table 2**
Side-by-side Comparison of Redshift Information Extracted by ML for Swift GRB Circulars (10 Randomly Selected Entries) with the Actual Redshift Information from the Swift GRB Table

| GCN | Predicted/Actual GRB | Predicted/Actual Redshift | Predicted/Actual Telescope | Predicted/Actual Redshift Type |
|---|---|---|---|---|
| 9712 | GRB 090726/090726 | 2.71/2.71 | SAO RAS 6 m telescope in Caucasus using SCORPIO/ SAO RAS | Spectroscopic/spectroscopic |
| 12867 | GRB 120119A/120119A | $z = 1.728/1.728$ | MMT 6.5 m telescope/MMT | Spectroscopic/spectroscopic |
| 4591 | GRB 060124/060124 | 0.82/2.30 | MDM 2.4 m telescope/MDM | Spectroscopic/spectroscopic |
| 18969 | GRB 160131A/160131A | 0.97/0.97 | Xinglong 2.16 m telescope/Xinglong | Spectroscopic/spectroscopic |
| 33110 | GRB 221226B/221226B | 2.694/2.694 | ESO VLT UT3 (Melipal)/VLT | Spectroscopic/spectroscopic |
| 7601 | GRB 080413B/080413B | 1.10/1.10 | ESO VLT/VLT | Spectroscopic/spectroscopic |
| 12761 | GRB 111228A/111228A | 0.716/0.716 | Gemini-South 8 m telescope/Gemini-South | Spectroscopic/spectroscopic |
| 7615 | GRB 080413B/080413B | 1.10/1.10 | Gemini-South/Gemini-South | Spectroscopic/spectroscopic |
| 11708 | GRB 110213A/110213A | $z = 1.46/1.46$ | 2.3 m Bok telescope/Bok | Spectroscopic/spectroscopic |
| 22039 | GRB 171020A/171020A | 1.87/1.87 | Nordic Optical Telescope (NOT)/NOT | Spectroscopic/spectroscopic |

**Note.** Leftmost column gives the GCN Circular number from which the LLM extracted the information from. Information generated by the LLM includes GRB numbers, redshift values or ranges, telescope names, and redshift types.

**Table 3**
Correct Predictions, Incorrect Predictions, and Overall Accuracy of Extracted Redshift Values on Positive, Negative, and All Examples in Our Evaluation Dataset

| | Positive | Negative | Total |
|---|---|---|---|
| Correct | 524 | 92 | 616 |
| Incorrect | 15 | 13 | 28 |
| Accuracy(%) | 97.2 | 87.6 | 95.7 |

**Table 4**
Correct Predictions, Incorrect Predictions, and Accuracies of Extracted Redshift Values, GRB Numbers, Telescope Names, and Redshift Types Evaluated against Positive Examples Taken from the Swift GRB Table

| | Redshift Value | GRB Number | Telescope Name | Redshift Type |
|---|---|---|---|---|
| Correct | 524 | 532 | 533 | 530 |
| Incorrect | 15 | 7 | 6 | 9 |
| Accuracy(%) | 97.2 | 98.7 | 98.9 | 98.3 |

number of correct and incorrect redshift predictions for each subset, along with the overall accuracy of our pipeline for redshift value extraction. Furthermore, we evaluated the model's ability to extract redshift values, GRB numbers, telescope names, and redshift types from the positive examples in our dataset. The results of this evaluation can be found in Table 4. We followed some general rules while evaluating the model's performance in each field. For the negative examples, any prediction that contained a numeric value was considered a false positive. For positive examples, any time the model predicted "No Redshift," we marked it as a false negative. If it predicted a wrong or bogus value for the redshift, it was marked as a false positive. In Circulars where multiple redshifts were reported, we would often mark the model's prediction as correct if it predicted any one of them, unless the Circular clearly mentioned that one of them was the most likely redshift for the GRB. Some other basic rules we followed during evaluation were:

1. For redshift ranges and limits, if the model extracted only the value of the upper or lower limit without indicating that it was a limit, we counted the redshift prediction as incorrect.
2. Predictions for GRB numbers had to match the actual GRB numbers mentioned in the Circular exactly; no trailing characters or letters were allowed.
3. We allowed some leeway in the case of telescope name predictions. Correct predictions of observatory or instrument names based on the information provided in the Circular were marked as correct, even if they did not match exactly with the telescope names reported in the Swift GRB table. For example, in GCN Circular 11195, our model predicted "Swift" as the telescope name, even though the Swift GRB table refers to it as "Swift-XRT." We still marked this prediction as correct, as it accurately identified the observatory name.

Finally, we evaluated the model's performance based solely on the information provided in each circular. There were instances where redshifts or GRB numbers in the Swift GRB table would be revised in later Circulars, causing the model's output to not match up with the data in the table. We marked our model's output as correct as long as its prediction was correct based on the information present within the circular in its prompt.

The most common errors made by the model for the negative examples dataset included:

1. *Hallucination of random numbers*. The model occasionally generated values that were not found in the original Circular, leading to inaccuracies. For example, in GCN Circular 10076, the model incorrectly predicted a redshift value of 0.342, which was not present in the Circular.
2. *Mistaking other numeric values*. Another common error was the misinterpretation of numeric figures for GRB redshift values. In GCN Circular 10046, the model extracted the reported power-law decay index of –0.95 as the redshift.

For the positive examples, the most frequent errors were:

1. *Reporting limits as exact values*. The model sometimes picked the lower or upper limits of redshift ranges as definite values. For example, in GCN Circular 12202, where the reported redshift range was $1.036 < z < 2.7$, the model returned just 1.036.

2. *False negatives in redshift*. There were instances where the model failed to report a redshift. For instance, GCN Circular 16891 reported tentative redshifts of 0.17 and 0.57 for GRB 141004A, which the model failed to extract.
3. *GRB number errors*. Occasionally, the model would get the GRB number correct but fail to include the letter, like in the case of GCN Circular 32099, where the model fails to add the letter "A" at the end of GRB 220521. These are marked as incorrect here.
4. *Telescope name errors*. Whenever the name of the telescope or observatory was not mentioned in the Circular, the model sometimes struggled to obtain it. For example, in Circular 5962, the telescope name "Magellan" was not provided in the text, and we still marked this as an error.

Despite all the above errors, the overall error rate remained relatively low, resulting in a high accuracy of 95.7% across the complete dataset. We recognize that further improvements in accuracy could be achieved with access to better hardware, which would allow us to run larger and more accurate versions of the model.

Upon manual evaluation of extracted data, we also found that some Circulars listed in the Swift GRB reference table as having reported redshifts did not actually contain redshifts. We counted five such Circulars in total. This may have been due to human error, which is often unavoidable. However, we noted our model was able to correct such instances four out of five times. Additionally, the model consistently predicted redshift types for the vast majority of Circulars, even when that information was not present in the Swift GRB table itself. This further underlines the potential of the information extraction pipeline: not only does it automate and speed up the process of astronomical data extraction, but it also improves it by compensating for human error and recovering missing classifications.

#### 3.2.2. Retrieval, Postprocessing, and Data Analysis

For the redshift extraction over the entire Circulars archive, a total of 1287 Circulars were retrieved using a combination of keyword and neural search methods. This accounted for approximately 3% of all Circulars in our database. Of these, 1095 Circulars were obtained through simple keyword matching in the subject line. An additional 192 Circulars were retrieved with the help of our neural search method. To evaluate the effectiveness of our retrieval methods, we calculated the number of redshift Circulars from our previous positive examples dataset that our method successfully retrieved. The keyword search alone helped us retrieve 505 out of the 539 redshift Circulars listed in the Swift GRB table. The neural search method contributed an additional 17 Circulars from this dataset, reducing the number of missing Circulars by half. Thus, in total, we retrieved 522 of the 539 Circulars in our positive examples dataset, giving us a retrieval recall rate of 96.8%. Manual analysis of the Circulars retrieved through neural search also shows that 175 of the additional 192 retrieved Circulars contained redshift values, ranges, limits, or estimates, giving a retrieval precision of approximately 91.1% for the neural search component of our methods. Thus, this method effectively extracted the majority of all redshift-reporting Circulars from our database.

Information was then extracted from all retrieved Circulars. On a single Colab L4 instance, the execution time was approximately 3.5 hr. From the total of 1287 Circulars, 1259 were parsed successfully. After data extraction, rows containing no redshifts were dropped, leaving us with 1078 entries in our redshift table.

After retrieving and generating output, the extracted redshift information required postprocessing before we could conduct data analysis. Our aim was to extract a single redshift value for each GRB in our output dataset. However, many Circulars contained redshift measurements for the same GRB events. Often, these would be reconfirmations of prior measurements, but sometimes would involve different measurements taken with different instruments or methods. To resolve these conflicting values, we used the following system. First, all identical GRB events in our dataset were aggregated. We then analyzed the redshift information extracted for these identical GRB events and resolved conflicts based on the following rules:

1. If both photometric and spectroscopic redshifts were reported, we prioritized the spectroscopic measurements.
2. If there were still multiple values, we selected the redshift with the highest reported decimal precision.
3. If conflicts still existed, we chose the redshift value that had the earliest reported date, using the "createdOn" field from the Circular `JSON`. This approach is based on the assumption that the earliest report is the original one, and later ones may be reconfirmations or reuse the previous values.

A total of 592 conflicting or multiple reported redshift values for the same GRB were removed. Among these 228 unique GRB events were extracted. Thus, after postprocessing, one Circular per GRB was selected and added back to our redshift table. Consequently, our final redshift table contained 714 unique GRB redshifts. Table 5 displays a randomly selected sample of 10 entries from this table.

Next, the redshift strings extracted from the Circulars were formatted into floating-point numbers for data visualization. Since the format of the redshift strings extracted from the Circulars was not always uniform, this step also required some text postprocessing. For example, GCN 10233 reported the redshift value with uncertainty as "0.49034 $\pm$ 0.00018." To extract a single redshift value per GRB, we applied the following rules:

1. Numeric redshift values were extracted as they were, without any associated uncertainties.
2. If only upper or lower limits are provided, we extracted the corresponding limits.
3. If a redshift range was provided, we determined its midpoint and used that as our redshift value.

Pattern matching with Python regular expressions was used to automate this step. Values greater than 8.2 were also removed, as the maximum reported redshift lies within this limit. In total, 687 of the 714 extracted redshift values were utilized for plotting.

Our analysis presents the extracted redshift values as a histogram in Figures 8 and 9, allowing us to learn the redshift distribution in GRBs utilizing ML techniques to derive the values from Circulars. We compare the frequencies of photometric and spectroscopic redshifts in Figure 8.

To study the behavior of the information extraction task using the LLM, we compile a table of the 10 highest redshifts

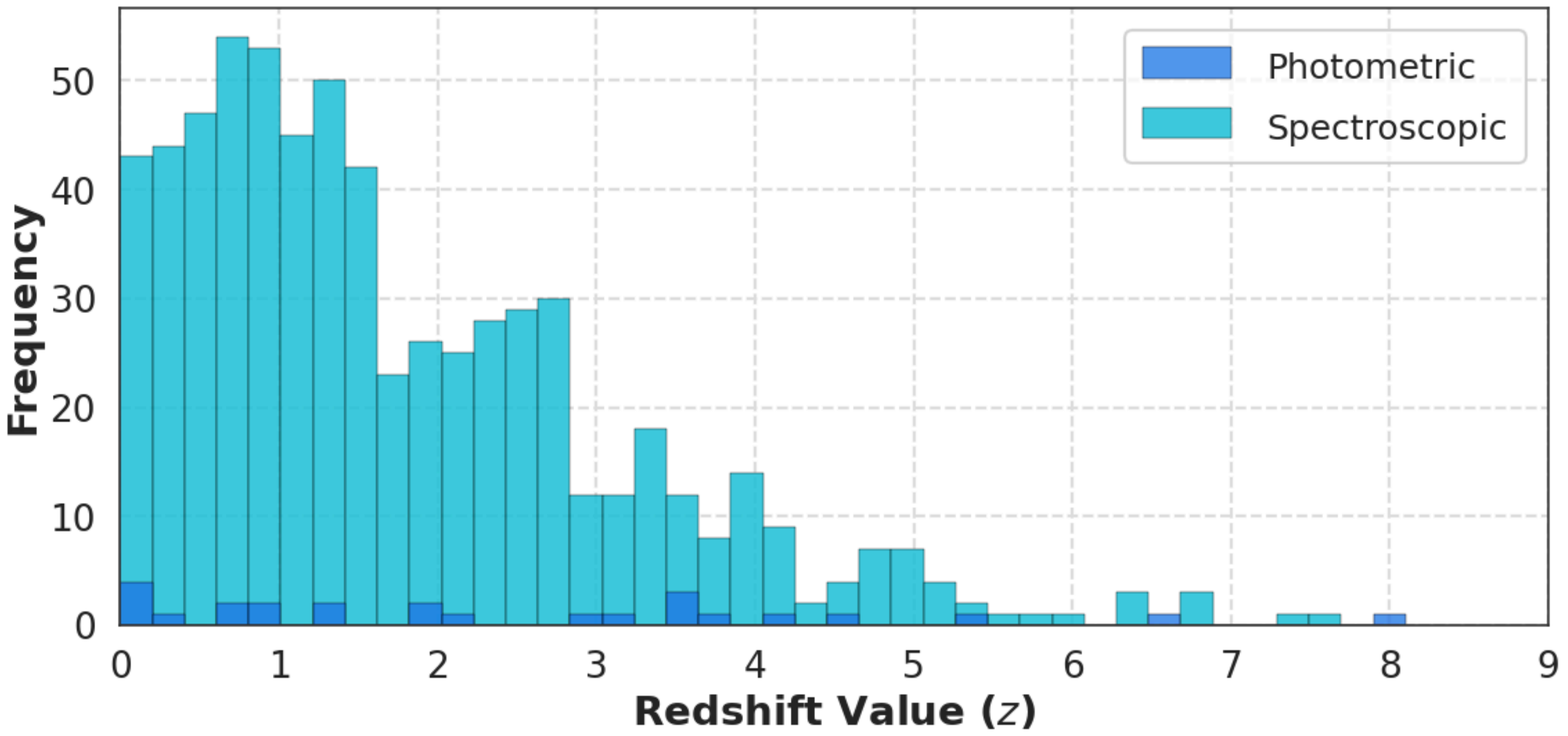

**Figure 8.** Photometric and spectroscopic redshift frequencies as reported in the GCN Circulars archive after postprocessing.

**Table 5**
Extracted Redshift Information (10 Randomly Selected Entries) from the Retrieved Circulars in the GCN Archive before Postprocessing

| GCN | GRB | Redshift | Telescope Name | Redshift Type |
|---|---|---|---|---|
| 16505 | GRB 140703A | $z = 3.14$ | GTC (Gran Telescopio Canarias) | Spectroscopic |
| 22096 | GRB 171010A | 0.3285 | ESO VLT | Spectroscopic |
| 10620 | GRB 100418A | 0.6235 | VLT/X-Shooter | Spectroscopic |
| 10441 | GRB 100219A | 4.8 | VLT | Spectroscopic |
| 8752 | GRB 081228 | $3.8 \pm 0.4$ | GROND | Photometric |
| 31107 | GRB 211023B | 0.862 | LBT-INAF | Spectroscopic |
| 3542 | GRB 050416(a) | $0.6535 \pm 0.0002$ | 10 m Keck I Telescope | Spectroscopic |
| 6952 | GRB 071020 | $z = 2.145$ | VLT | Spectroscopic |
| 17758 | GRB 150424A | $z = 0.30$ | 10.4 m GTC telescope (+OSIRIS) | Spectroscopic |
| 38097 | GRB 241105A | 2.681 | ESO VLT UT1 (Antu) | Spectroscopic |

**Note.** GCN numbers are taken from the "CircularId" field of the Circulars, while predicted redshifts, GRB numbers, telescope names, and redshift types are all generated by `Mistral 7B Instruct` after prompt tuning and output parsing with `LangChain`. (Note: The blank space between 3.3721 and 0.0004 for the redshift of GCN 9015 should be interpreted as a ± sign. The actual redshift is thus 3.3721 ± 0.0004. This happens due to the corruption of certain characters during text preprocessing. Also note that the model makes an error for the GRB prediction of GCN 32099, extracting a GCN number from the circular instead.)

reported in the Circulars, as shown in Table 6. As a caveat, we note that the LLM does not distinguish between observationally confirmed values and those inferred from model assumptions or values of theoretical considerations mentioned in the Circulars. Since the highest spectroscopically confirmed redshift, GRB 090423, is at $z \sim 8.1$ (A. Fernandez-Soto et al. 2009), we truncated values up to that limit. Without this truncation, the LLM would additionally extract redshifts reported only as theoretical predictions or assumed for model parameters. Additionally, we note that the ML model does not provide accurate classifications for the type of redshift; for example, GRB 090423 and GRB 060116 have spectroscopic redshifts but are misclassified as photometric by the model.

## 4. Discussion and Conclusion

Our study highlights the potential of modern NLP techniques and open-source LLMs to extract insights from large astrophysical text databases. We successfully applied neural topic modeling techniques to uncover common astrophysical themes and trends present within the rich GCN Circular archive. By employing contrastive fine-tuning techniques on transformer-based language models, we classified the Circulars into five categories based on the type of observations reported, allowing us to collect some frequency statistics and examine trends for various observation types in GCN. Additionally, we specifically classified GW Circulars, their counterparts, and non-GW Circulars, and we observed a clear increase in interest in multimessenger astronomy since 2015. Furthermore, we implemented a zero-shot pipeline using an open-source language model to retrieve and extract GRB redshift information from the GCN archive. When evaluated on redshift-containing Circulars from the Swift observatory's GRB table, our pipeline achieved an accuracy of over 97% on redshift value/range prediction and over 98% on event names, telescope names, and redshift types prediction. Additionally, by integrating RAG techniques using a combination of keyword and neural search, our pipeline was able to extract a substantial amount of GRB redshift data automatically. This significantly reduced the need for manual data extraction. Our study indicates that expensive training and private models are not necessary to achieve the results in astrophysical information extraction. With simple prompt modifications, it is possible to extract a wide variety of desired information for further data analysis.

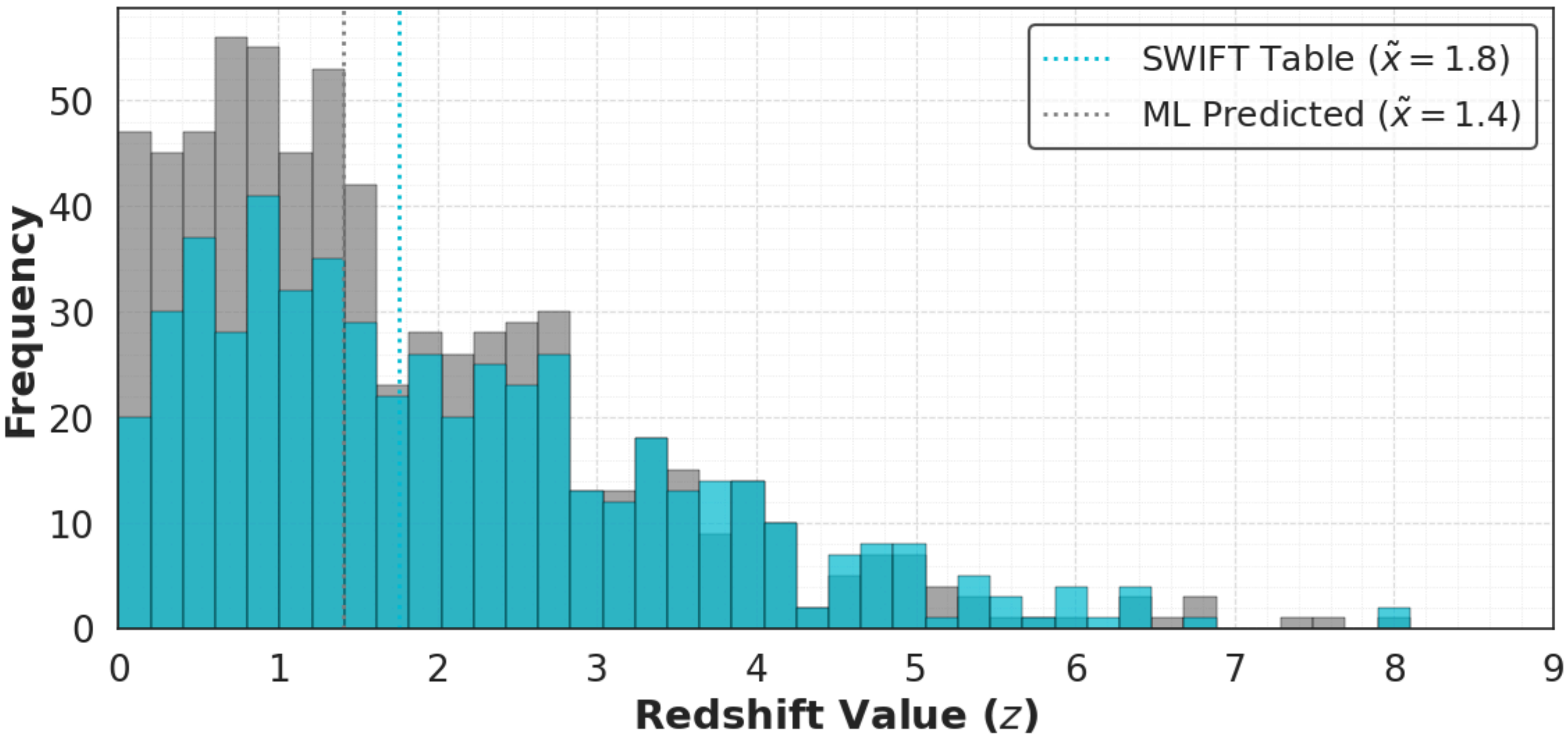

**Figure 9.** Distribution of Swift GRB redshifts (cyan) vs. ML-predicted postprocessed redshifts (gray), along with their median values.

**Table 6**
Highest Redshift Values Extracted, after Removal of Errors, before Postprocessing

| GCN | GRB | Redshift | Telescope Name | Redshift Type |
|---|---|---|---|---|
| 9222 | GRB 090423 | $z \sim 8.1$ | NICS/Amici combination | Photometric |
| 9216 | GRB 090423 | $z \sim 7.6$ | Italian TNG 3.6 m telescope | Spectroscopic |
| 39732 | GRB 250314A | 7.3 | ESO VLT | Spectroscopic |
| 35756 | GRB 240218A | 6.782 | ESO/VLT UT3 (Melipal) | Spectroscopic |
| 8225 | GRB 080913 | $z = 6.7$ | VLT (Very Large Telescope) | Spectroscopic |
| 8256 | GRB 080913 | z = 6.7 | Swift-BAT and Konus-Wind | Spectroscopic |
| 4545 | GRB 060116 | 6.6 | ESO VLT | Photometric |
| 16269 | GRB 140515A | $z = 6.32$ | GMOS on Gemini-N 8 m telescope | Spectroscopic |
| 30771 | GRB 210905A | 6.318 | ESO VLT Melipal | Spectroscopic |
| 3937 | GRB 050904 | 6.29 | Subaru 8.2 m telescope | Spectroscopic |

Structured extraction of information in standardized formats such as .txt or .json can greatly streamline the construction of light curves and spectral energy distributions, enabling rapid modeling of afterglow behavior. With access to this crucial information, astronomers can then make near-real-time decisions about observation scheduling, calculate required exposure times based on decay slopes, and coordinate multiwavelength campaigns far more efficiently. Thus, in the future, by broadening the scope from redshift-centric analytics to a multiparameter, filter-based extraction system, as a real-time pipeline by leveraging the LLM-powered approach, is transformative for designing a follow-up-ready intelligence platform. This approach would directly benefit the global GRB observation network, ensuring that crucial observational opportunities are not lost in the hours following the burst.

We anticipate LLMs will be useful for enhancing the utility of the Circulars archive and even the astronomical alerts database, particularly by providing rapid and precise access to key observational parameters to the follow-up community. While the current pipeline only extracts redshift data, an LLM-based system could be trained to parse and standardize to a much broader range of critical metadata, such as time since burst ($T - T_0$), exposure times, filters, magnitudes, magnitude systems (AB or Vega), telescope/instrument names, and the Circular number itself. By recognizing and classifying the photometric observations based on UV, optical, and NIR filter information, the model may enable instantaneous, filter-specific data sorting and retrieval, which is currently limited by the need for manual processing. Such a capability would significantly enhance the scientific readiness for the follow-up community, particularly for optical/NIR/radio afterglow observations.

Automatic prompt and query generation through the use of autonomous agents is an upcoming area of research within the NLP space, which could potentially help improve the accuracy and robustness of our information extraction pipeline (S. Barua 2024). Additionally, the testing of our pipeline with larger LLMs can help us realize the true potential of state-of-the-art NLP methods for astrophysical data extraction. In the near future, we aim to investigate how our NLP pipelines can be integrated into the GCN platform, potentially functioning as an AI assistant for astronomy.

### 4.1. Challenges and Limitations

In our research, there are some limitations. The observations-based classifications require the manual collection of data to fine-tune our transformer model. This means that if we wanted to classify based on different criteria, we would need to collect new training datasets accordingly. For our information extraction pipeline, the lacked access to high-end GPUs forced us to rely on those provided by Google Colab. As a result, we had to use smaller LLM models that had more limited accuracy compared to larger models, as these were the only ones that could fit into the available GPUs. We thus believe that the accuracy of our information extraction pipeline could be improved considerably through the use of better hardware. Furthermore, our pipeline required us to use some manual output parsing techniques in addition to automatic ones to successfully parse the vast majority of Circulars. Occasional nonuniformity of the model output was the cause for this, which could have been reduced with larger, more accurate instruction-tuned LLMs, along with more refined prompt tuning strategies. The prompt tuning itself was an iterative process that required domain knowledge and several attempts to achieve satisfactory results. Finally, despite our best efforts at reducing hallucination through the use of RAG, a small number of errors still slipped in, partly due to the retrieval of Circulars that did not contain redshift information. Depending upon the required accuracy of the information retrieval task at hand, this may make the pipeline in its current state not suitable for highly sensitive and accurate tasks.

As a caveat, we also acknowledge that NLP is a rapidly evolving field, with new LLM models and libraries emerging at a fast pace, which presents both limitations and opportunities. In this study, we employed BERTopic and `Mistral` models for topic modeling and information extraction, relying on model architectures and tools developed in recent years. While these models have produced promising initial results, we recognize that future developments may offer improved performance and better accuracy. This first application of BERTopic and `Mistral` models for a multimessenger alert system provides a foundation upon which future work can build.

## Acknowledgments

We thank the anonymous referee for useful comments and suggestions on the manuscript. V.S. was sponsored by support from NASA through a cooperative agreement with the Center for Research and Exploration in Space Science and Technology II (CRESST II). The views and conclusions contained in this document are those of the authors and should not be interpreted as representing the official policies, either expressed or implied, of NASA or the US Government. The US Government is authorized to reproduce and distribute reprints for Government purposes, notwithstanding any copyright notation herein. The GCN team acknowledges support from NASA's Internal Scientist Funding Model (ISFM) program. This research has made use of data obtained through the GCN Service, provided by the NASA Goddard Space Flight Center (GSFC), in support of NASA's High Energy Astrophysics Programs. The authors would also like to thank Daniela Huppenkothen for the insightful discussions. R.G. was sponsored by NASA through a contract with ORAU. M.W.C. acknowledges support from the National Science Foundation (NSF) under grant Nos. PHY-2409481, PHY-2308862, and PHY-2117997. N.M. acknowledges support from NSF under awards PHY-1764464 and PHY-2309200 to the LIGO Laboratory, under Cooperative Agreement PHY-2019786 (The NSF AI Institute for Artificial Intelligence and Fundamental Interactions, http://iaifi.org/), and from MathWorks, Inc.

## Code Availability

All codes are available to the wider astrophysical community via Zenodo (doi:10.5281/zenodo.17538208) and the GitHub public repository[19] at `nasa-gcn`. The repositories contain data, figures, and code in the Google Colab notebooks and extended tables.

## Appendix A
## Word Cloud Representation of GCN Circulars

Figure 10 shows a word cloud generated over the entire GCN Circulars archive.

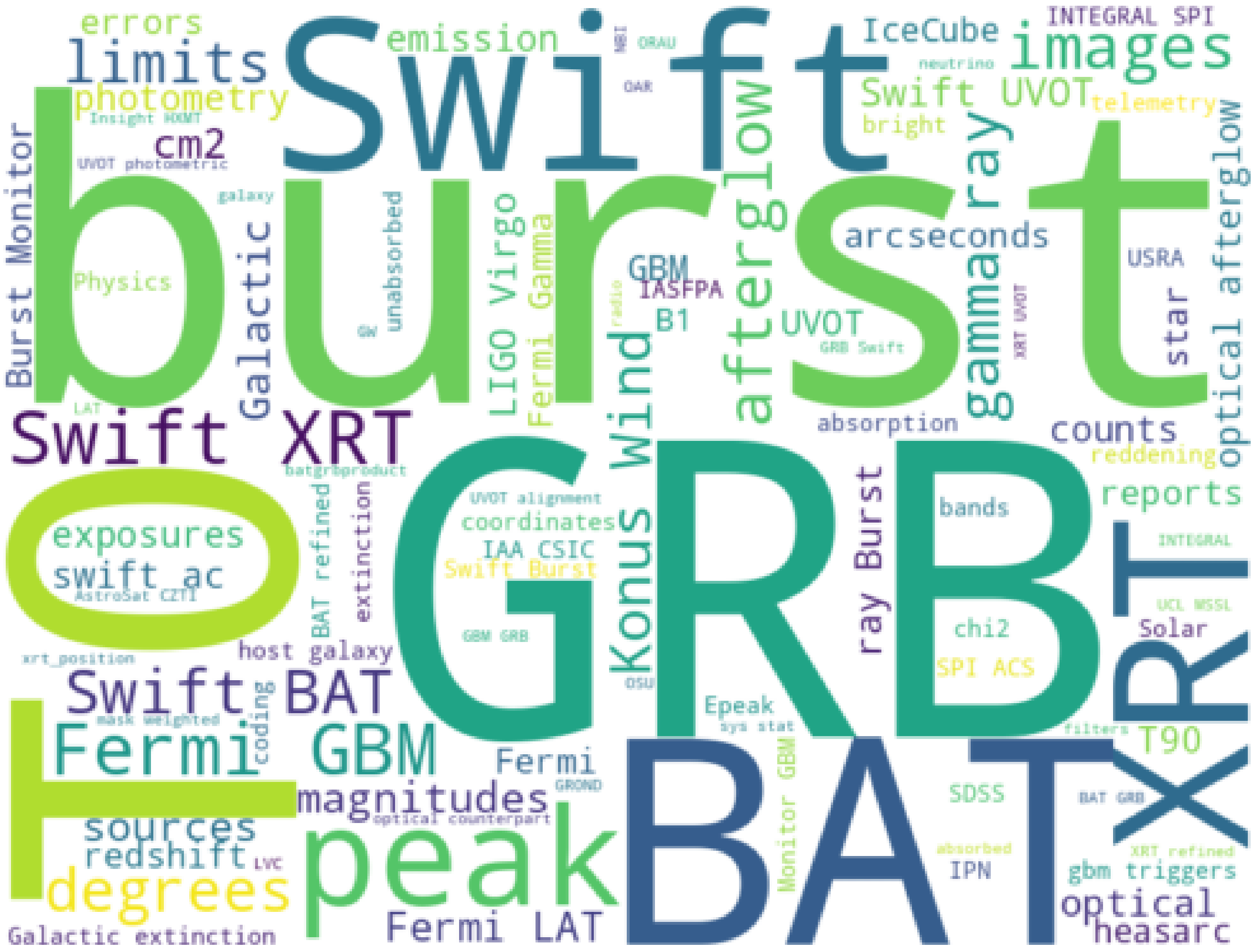

**Figure 10.** GCN Word Cloud.

## Appendix B
## Observations and GW Similarity Matrix

Figure 11(a) shows the similarity matrix of all five observation types, created using BERTopic. The topic embeddings of each observation-based cluster are used to create this cosine similarity matrix.

Figure 11(b) presents the gravitational event map generated using BERTopic. The topic embeddings of each gravitational-based cluster are used to create this cosine similarity matrix.

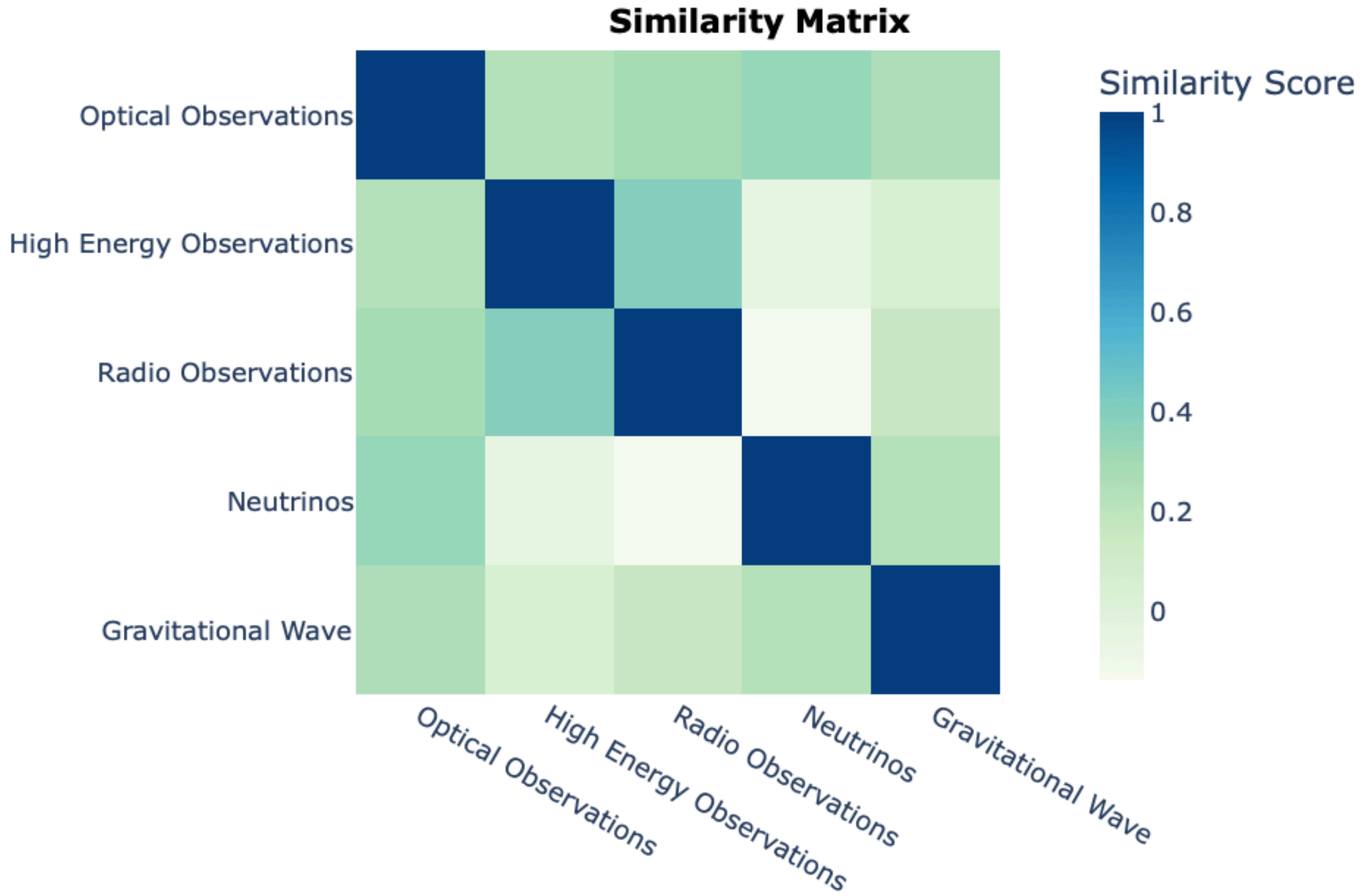

**(a)** Observation-based clusters.

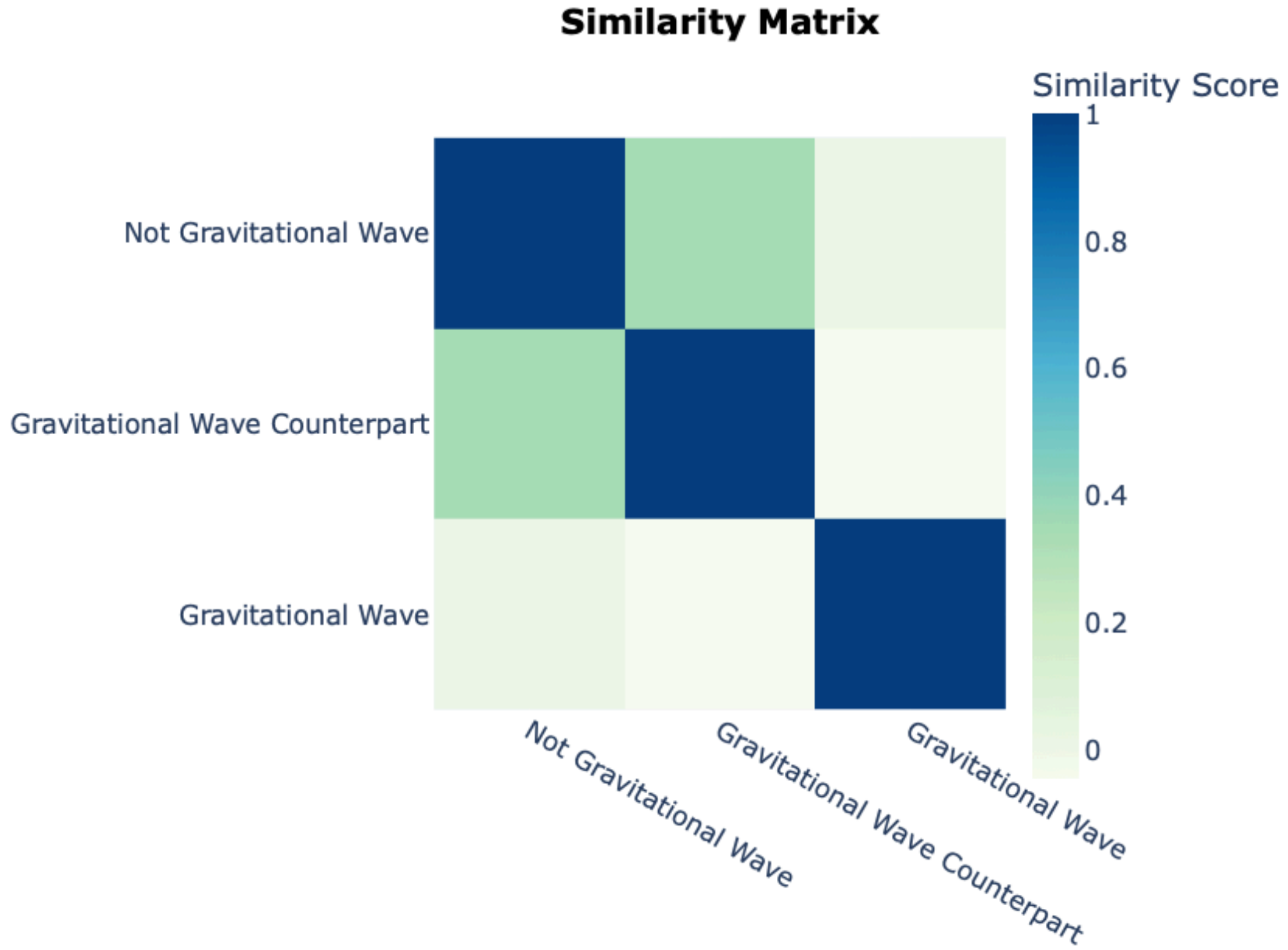

**(b)** Gravitational-wave focused clusters.

**Figure 11.** Similarity matrices for (a) all five observation types and (b) GW-focused clusters. Each type is represented with its most important keywords extracted using TF-IDF.

## Appendix C
## Retrieved and Extracted Redshifts Table

Finally, all retrieved Circulars with their extracted redshift information, before postprocessing, are shown in Table 7.

**Table 7**
Retrieved and Extracted Redshifts

| GCN | GRB | Redshift | Telescope Name | Redshift Type |
|---|---|---|---|---|
| 12 | GRB 971214 | 1.02 | Keck II 10-m telescope | Spectroscopic |
| 13 | GRB 971214 | 1.03 | Keck II 10-m telescope | Spectroscopic |
| 29 | GRB 971214 | $z = 3.43$ | Keck II telescope | Spectroscopic |
| 137 | GRB 980703 | 0.967 | Keck II telescope | Spectroscopic |
| 139 | GRB 980703 | $0.9660 \pm 0.0002$ | Caltech GRB Collaboration | Spectroscopic |
| 189 | GRB 980613 | $1.0964 \pm 0.0003$ | Keck I | Spectroscopic |
| 216 | GRB 990123 | Approximately 0.2–0.3 | The Caltech-CARA-NRAO Collaboration | Photometric |
| 219 | GRB 990123 | $1.600 \pm 0.001$ | NOT, Copenhagen, IAC, LAEFF, IAA | Spectroscopic |
| 223 | GRB 990123 | $z \sim 0.2$ | Keck | Spectroscopic |
| 249 | GRB 990123 | $1.598 \pm 0.001$ | NOT, Copenhagen, IAC, IAA, LAEFF, IAA | Spectroscopic |

(This table is available in its entirety in machine-readable form in the online article.)

## ORCID iDs

Vidushi Sharma https://orcid.org/0000-0002-4394-4138
Ronit Agarwala https://orcid.org/0009-0004-9222-0402
Judith L. Racusin https://orcid.org/0000-0002-4744-9898
Leo P. Singer https://orcid.org/0000-0001-9898-5597
Tyler Barna https://orcid.org/0000-0002-4843-345X
Eric Burns https://orcid.org/0000-0002-2942-3379
Michael W. Coughlin https://orcid.org/0000-0002-8262-2924
Dakota Dutko https://orcid.org/0009-0007-6220-2192
Courey Elliott https://orcid.org/0009-0000-1860-9509
Rahul Gupta https://orcid.org/0000-0003-4905-7801
Ashish Mahabal https://orcid.org/0000-0003-2242-0244
Nikhil Mukund https://orcid.org/0000-0002-8666-9156